\documentclass[11pt]{article}
\usepackage[utf8]{inputenc}
\usepackage{geometry}
\usepackage{titling}
\usepackage{authblk}
\usepackage{amssymb}
\usepackage{amsmath}
\usepackage{amsthm}
\allowdisplaybreaks[4]
\usepackage{slashed}
\usepackage{graphicx,color,epsfig}
\usepackage{cite}
\usepackage[colorlinks=true,linkcolor=red,citecolor=blue]{hyperref}
\begin{document}
\title{Resonant Contributions to Three-body $B\to KKK$ Decays in Perturbative QCD Approach}
\author[1]{Zhi-Tian Zou$\footnote{zouzt@ytu.edu.cn}$}
\author[1,2]{Ying Li$\footnote{liying@ytu.edu.cn}$}
\author[1]{Qi-Xin Li}
\author[3]{Xin Liu$\footnote{liuxin@jsnu.edu.cn}$}
\affil[1]{\it Department of Physics, Yantai University, Yantai 264005, China}
\affil[2]{\it Center for High Energy Physics, Peking University, Beijing 100871, China}
\affil[3]{\it Department of Physics, Jiangsu Normal University, Xuzhou 221116, China}
\maketitle
\vspace{0.2cm}

\begin{abstract}
In this work, we study the ($S$, $P$ and $D$)-wave  $K^+K^-$ contributions to $B\to KKK$ decays in the perturbative QCD approach at leading order. Within the two-meson wave functions describing the nonperturbative dynamics in the kaon-pair for different waves, we calculate the branching fractions and the direct $CP$ asymmetries of these decay modes in the corresponding resonance regions.  Most of our numerical results are well consistent with the current measurements. We note that the narrow-width approximation is invalid in the quasi-two-body decays $B\to Kf_0(980)\to KKK$. For other decays, under the narrow-width approximation we can extract the branching fractions of the corresponding two-body decays involving the intermediate resonant states, and the related branching fractions agree with the current experimental data well. Furthermore, we also predict the corresponding quasi-two-body decays $B\to K\pi^+\pi^-$, which are expected to be measured in the ongoing LHCb and Belle-II experiments.
\end{abstract}

\section{Introduction}
Studies of $B$ meson decays to three-body charmless hadronic final states are a natural extension of studies of decays to two-body charmless final states. Some of the final states considered so far as two-body (for example $\phi K$, $f_0K$, etc.) proceed via quasi-two-body processes involving a wide resonance state that immediately decays in the simplest case to two particles, thereby producing a three-body final state. Multiple resonances occurring nearby in phase space will interfere and a full amplitude analysis is required to extract correct branching fractions for the intermediate quasi-two-body states. In past few years, more and more analysis of three-body decays have been performed by the BaBar \cite{Aubert:2003mi, Aubert:2005ce, Aubert:2006nu, Aubert:2009av,  Aubert:2007sd, Aubert:2007bs, Aubert:2008bj, Aubert:2009me, Lees:2011nf, BABAR:2011ae, Lees:2012kxa}, Belle \cite{Abe:2002av, Garmash:2003er, Garmash:2004wa, Garmash:2005rv, Garmash:2006fh, Dalseno:2008wwa, Nakahama:2010nj}, CLEO \cite{Eckhart:2002qr} and LHCb \cite{Aaij:2013sfa, Aaij:2013bla, Aaij:2014iva, Aaij:2016qnm, Aaij:2018rol, Aaij:2019nmr, Aaij:2017zgz, Aaij:2013orb}, and the branching fractions and CP violations have been measured with high precision, which could provide us possibilities  for testing the standard model (SM), exploring the source of $CP$ violation and searching for the possible effects from new physics (NP) beyond SM \cite{Cheng:2009xz,Li:2018lxi}. For example, in the $B_s^0\to K_S^0K^{\pm}\pi^{\mp}$ decays, the final states $K_S^0K^-\pi^+$ and $K_S^0K^+\pi^-$ are not flavor-specific, both $B_s$ and $\overline{B}_s$ can decay to these two modes, with the corresponding amplitudes excepted to be comparable in magnitude. The large interference shall lead to the large $CP$ asymmetries, providing us new possibilities for $CP$ violation searches. As we known, some tree-level open-charm $B$ decays are theoretically clean to determine the angle $\gamma$ of the Cabibbo-Kobayashi-Maskawa (CKM) unitarity triangle, such as the $B_s\to \overline{D}^0\phi$ decay. Because in the experiments $\phi$ meson is  reconstructed within $K^+K^-$ final states, so the analysis of corresponding three-body decay $B_s\to \overline{D}^0K^+K^-$ could further improve the determination of $\gamma$. Also, within this decay, the small phase $\phi_s$ in $B_s^0-\overline{B}_s^0$ mixing can be well determined with as small theoretical uncertainties as possible \cite{Nandi:2011uw}. Some decays such as $B^0 \to K_SK_SK_S$ mediated by the flavour-changing neutral-current $b\to s$ transition provide a sensitive probe of the effect of new physics beyond SM. Motivated by the experimental results, many theoretical studies of various three-body non-leptonic $B$ decays have been performed in different frameworks, such as approaches based on the symmetry principles \cite{Gronau:2005ax, Engelhard:2005hu, Imbeault:2011jz, Bhattacharya:2013boa, He:2014xha}, the QCD factorization (QCDF) \cite{ElBennich:2009da, Krankl:2015fha, Cheng:2002qu, Cheng:2007si, Cheng:2016shb, Cheng:2014uga, Cheng:2013dua, Li:2014oca, Li:2014fla}, the perturbative QCD approach (PQCD) \cite{Wang:2014ira, Wang:2015uea, Wang:2016rlo, Li:2016tpn, Wang:2017hao, Ma:2019sjo, Li:2018qrm, Li:2018lbd, Ma:2017aie, Li:2017obb, Ma:2017kec, Li:2017mao, Ma:2016csn, Li:2015tja, Ma:2017idu, Rui:2017bgg, Rui:2017fje, Rui:2018hls, Rui:2019yxx, Li:2019pzx, Li:2019hnt, Rui:2017hks, Xing:2019xti, Cui:2019khu}, and other theoretical methods \cite{Zhang:2013oqa, Wang:2015ula, Qi:2018syl, ElBennich:2006yi}. {

In comparison with two-body decays, $B$ meson hadronic three-body decays are much more complicated, because they receive contributions not only from resonance and nonresonance, but also from the possible final state interactions among the final particles. The relative strengths of these contributions vary remarkably for different modes. Based on the well measured branching fractions from the resonant and nonresonant components \cite{ Aubert:2008bj, Aubert:2009me, Lees:2011nf, BABAR:2011ae,Lees:2012kxa, Garmash:2004wa, Garmash:2005rv, Aaij:2016qnm, Aaij:2018rol}, it is found that the nonresonant contributions play essential roles in penguin dominant three-body $B$ decays. For example, the nonresonant fractions can be as large as $(70-90)\%$ in $B\to KKK$ decays, while in  $B\to\pi\pi\pi$ decays that are induced by the tree diagrams the nonresonant fractions are as small as $40\%$. Moreover, for the weak $B$ decays, the release energy is of order $5 \rm GeV$ and most resonances lie in the region of $0.5 \sim 2$ GeV, so it is possible to get sizable nonresonant contributions from three-body charmless $B$ decays. In this sense, the explicit theoretical studies will help us to disentangle the resonant and nonresonant contributions, and further improve the understanding of the unclear nonresonant mechanism.

As aforementioned, some of the final states proceed via quasi-two-body processes and many resonances are involved. So far, all attempts to interpret the effects of the resonances are still model dependent, such as the isobar model \cite{Sternheimer:1961zz, Herndon:1973yn} and the K-matrix formalism \cite{Chung:1995dx}. The Dalitz plot analysis allow one to investigate the resonant contributions within the isobar model, which is popularly applied to describe the complex decay amplitude by experimentalists. In the configuration of the quasi-two-body process, the two energetic particles produced from the inner resonance are collinear and form a moving-fast meson-pair, then the interactions between the meson-pair and the bachelor particle are power suppressed naturally. The interaction in the meson-pair can be described by the two-meson wave function. In this picture, in such quasi-two-body region of phase space, the obvious generalization of the factorization theorem for two-body decays applies.  It is reasonable for us to assume the validity of the factorization for these quasi two-body $B$ decays. Based on the argued factorization and using the two-meson wave function, in the PQCD framework that is based on the $k_{\rm T}$ factorization, the decay amplitude of quasi-two-body $B$ decays can be decomposed as the convolution
\begin{equation}\label{convention}
\mathcal{A}=\Phi_B\otimes\mathcal{H}\otimes\Phi_{M_1M_2}\otimes\Phi_{M_3},
\end{equation}
where the $\Phi_{B}$, $\Phi_{M_3}$ are the wave functions of $B$ meson and the light bachelor meson, respectively.  $\Phi_{M_1M_2}$ is the two-meson wave function in resonant region. The hard kernel $\mathcal{H}$ for the $b$ quark decay, similar to the two-body case, starts with the diagrams of single hard gluon exchange. An advantage of the above formalism is that both resonant and nonresonant contributions to the hadron-pair system can be included into the wave function through appropriate parametrization.

In this work, we shall focus on the $B\to KKK$ decays dominated by the flavor-changing neutral-current $b\to s$ transitions, which are sensitive to NP beyond SM. In ref.\cite{Garmash:2004wa}, based on a 140 $fb^{-1}$ data sample containing $152\times10^6$ $B\overline{B}$ pairs, Belle collaboration performed a full amplitude analysis to the $B^+\to K^+K^+K^-$ decay for the first time, and found that there are two obvious peaks in the two-particle invariant mass spectra. One is a narrow peak at 1.02 GeV corresponding to the $\phi(1020)$ meson, while another a broad structure around 1.5 GeV, which was referred to as $f_X(1500)$. In 2012, BaBar collaboration also improved their measurements and performed a detailed analysis for the $B^+\to K^+K^+K^-$ and $B^0\to K^0K^+K^-$ decays \cite{Lees:2011nf, Lees:2012kxa}, based on a data sample of approximately $470\times 10^6$ $B\overline{B}$ decays. The large peak around $m(K^+K^-)\sim$ 1.5 GeV was also observed. Because the interpretation of the $f_X(1500)$ state is uncertain, both Belle and BaBar have modeled it as a scalar resonance, though a vector structure cannot be ruled out. In $B^+\to \pi^+K^+K^-$ decay \cite{Aubert:2007xb}, BaBar collaboration also reported a peak around $m(K^+K^-)\sim$ 1.5 GeV, but they did not find the obvious evidences of $f_X(1500)$ in decays $B^{\pm}\to \pi^{\pm}K_SK_S$ \cite{Aubert:2008aw} and $B^0\to K_SK_SK_S$\cite{Lees:2011nf}. To identify physical properties and quantum numbers of the $f_X(1500)$, larger data samples are needed, especially the measurements of the decays involving $K_SK_S$ pair, because only even spin resonances can decay to $K_S K_S$ final states, according to the Bose-Einstein statistics. If $f_X(1500)\to K_S K_S$ were observed experimentally, we then could confirm that $f_X(1500)$ is an even-spin structure.

In recent years, within PQCD approach, the quasi-two-body $B$ meson decays including $\pi\pi$ pair and $K\pi$ pair through the $S$, $P$, and $D$ wave resonances have been studied extensively \cite{Wang:2016rlo, Li:2016tpn, Ma:2019sjo, Wang:2017hao, Li:2018qrm, Li:2018lbd, Ma:2017aie, Li:2017obb, Ma:2017kec, Li:2017mao, Ma:2017idu, Ma:2016csn, Li:2015tja}. In ref.\cite{Rui:2019yxx}, the authors have studied the $B_s$ decays to charmonium and $K\bar K$-pair, motivated by the LHCb measurements\cite{Aaij:2017zgz,Aaij:2013orb}. In this article, we restrict ourselves to these three-body $B$ decays involving three kaons in final states with accounting for the $S$, $P$, and $D$ wave resonant contributions, stimulated by the Belle and BaBar measurements \cite{Garmash:2004wa,Lees:2012kxa,Lees:2011nf}. Besides, we will account for the following resonances, $f_0(980)$, $f_0(1500)$, $f_0(1710)$, $\phi(1020)$, $f_2^\prime(1525)$, and $f_2(2010)$, which have been detailed analyzed in ref.\cite{Lees:2012kxa, Lees:2011nf} using the Dalitz plot in the experiments.  In technical aspect, we shall also follow the PQCD framework of quasi-two-body mechanism to investigate the resonant contributions in detail. For the $CP$ asymmetries, we shall only discuss the direct $CP$ asymmetry, leaving the $CP$ violations induced by the interference between the intermediate resonances for the future.

The outline of the present paper is as follows. In Sec.\ref{sec:intro}, we firstly introduce the formalism of PQCD on three-body of $B$ decays, and the decay formalism will be given. The detailed analytic calculations will be presented in Sec.\ref{sec:amplitude}. In Sec.\ref{sec:result}, we will address the numerical results, including the branching fractions and the localized $CP$ asymmetries. Combining the experimental data and the obtained theoretical results, we also perform the discussions in this section. Finally, we will summarize our work in Sec.\ref{sec:summary}

\section{Framework}\label{sec:intro}
In the quasi-two-body region of phase space, the Dalitz plot analysis allows one to describe the decay amplitude in the isobar model, where the decay amplitude is represented by a coherent sum of amplitudes from  $N$ individual decay channels with different resonances,
\begin{eqnarray} \label{isobar}
\mathcal{A}=\sum_{j=1}^{N} a_j \mathcal{A}_j,
\end{eqnarray}
where the $\mathcal{A}_j$ is the amplitude corresponding to certain resonance and $a_j$ is the complex coefficient describing the relevant magnitude and phase of the different decay channel. From this equation, one can easily find that there exist not only the direct $CP$ asymmetry for particular intermediate resonance but also the $CP$ asymmetries induced by the interferences among different resonances.

For the penguin dominant $B_{u,d}\to KKK$ decays, the weak Hamiltonian $\mathcal{H}_{eff}$ of $b\to s q \bar q$ can be decomposed as \cite{Buchalla:1995vs}
\begin{eqnarray}
\mathcal{H}_{eff}=\frac{G_F}{\sqrt{2}}\Big\{V^*_{ub}V_{us}(C_1O_1+C_2O_2)-V^*_{tb}V_{ts}\sum_{i=3}^{10}C_i O_i\Big\},
\end{eqnarray}
where the $V_i$ are the CKM matrix elements. The $C_{i}(i=1,...,10)$ is the Wilson coefficient corresponding to the four-quark operator $O_{i}$. The tree operators $O_{1,2}$ are written as
\begin{eqnarray}
O_1=(\bar{b}_{\alpha}u_{\beta})_{V-A}(\bar{u}_{\beta}s_{\alpha})_{V-A},
O_2=(\bar{b}_{\alpha}u_{\alpha})_{V-A}(\bar{u}_{\beta}s_{\beta})_{V-A},
\end{eqnarray}
where $\alpha$ and $\beta$ are the color indexes. For the QCD and electroweak penguin operators, the explicit expressions are listed as
\begin{eqnarray}
&&O_3=(\bar{b}_{\alpha}s_{\alpha})_{V-A}\sum_{q=u,d,s}(\bar{q}_{\beta}q_{\beta})_{V-A},\;
O_4=(\bar{b}_{\alpha}s_{\beta})_{V-A}\sum_{q=u,d,s}(\bar{q}_{\beta}q_{\alpha})_{V-A},\\
&&O_5=(\bar{b}_{\alpha}s_{\alpha})_{V-A}\sum_{q=u,d,s}(\bar{q}_{\beta}q_{\beta})_{V+A},\;
O_6=(\bar{b}_{\alpha}s_{\beta})_{V-A}\sum_{q=u,d,s}(\bar{q}_{\beta}q_{\alpha})_{V+A},\\
&&O_7=\frac{3}{2}(\bar{b}_{\alpha}s_{\alpha})_{V-A}\sum_{q=u,d,s}e_q(\bar{q}_{\beta}q_{\beta})_{V+A},\;
O_8=\frac{3}{2}(\bar{b}_{\alpha}s_{\beta})_{V-A}\sum_{q=u,d,s}e_q(\bar{q}_{\beta}q_{\alpha})_{V+A},\\
&&O_9=\frac{3}{2}(\bar{b}_{\alpha}s_{\alpha})_{V-A}\sum_{q=u,d,s}e_q(\bar{q}_{\beta}q_{\beta})_{V-A},\;
O_{10}=\frac{3}{2}(\bar{b}_{\alpha}s_{\beta})_{V-A}\sum_{q=u,d,s}e_q(\bar{q}_{\beta}q_{\alpha})_{V-A},
\end{eqnarray}
where the $e_q$ is the charge of the active quark $q$.

In Eq.(\ref{convention}), the key step in the theoretical studies is how to describe the nonperturbative parts properly reflected by the wave functions, as they are the most important inputs in PQCD approach. The wave functions of the $B$ meson and the $K$ meson have been well determined by those well measured charmless/charmed two-body $B$ decays in experiments, such as $B \to KK, K\pi, DK$ decays \cite{Keum:2000ph, Lu:2000em,Yu:2005rh, Ali:2007ff}, and we are not going to discuss them any more in this paper. Compared to the $B$ meson two-body decays, in the quasi-two-body decays the new ingredient is the two-meson wave functions corresponding to different resonances with different spin.

We first discuss the $S$-wave two-meson wave function of the $K\bar K$-pair \cite{Rui:2019yxx}, whose form is the same as the $\pi\pi$ pair and can be written as \cite{Wang:2015uea, Xing:2019xti}:
\begin{eqnarray}
\Phi_S=\frac{1}{\sqrt{2Nc}}[P\mkern-11.5mu/\phi_S(z,\xi,\omega^2)+\omega\phi_S^s(z,\xi,\omega^2)
+\omega(n\mkern-9.5mu/ v\mkern-7.5mu/-1)\phi_S^t(z,\xi,\omega^2)],
\end{eqnarray}
where $z$ is the momentum fraction of the spectator quark, and $\xi$ is the momentum fraction of one $K$ in the $K\bar K$-pair. $\omega$ and $P$ are the invariant mass and momentum of the $K\bar K$-pair, respectively. $n=(1, 0, 0_{\rm T} )$ and $v=(0, 1, 0_{\rm T} )$ are two dimensionless vectors. The $\phi_S$, $\phi_S^s$, $\phi_S^t$ are the twist-2 and twist-3 distribution amplitudes, and they are parameterized as \cite{Wang:2015uea,Diehl:1998dk}
\begin{eqnarray}
\phi_S(z,\xi,\omega^2)&=&\frac{9}{\sqrt{2Nc}}F_S(\omega^2)a_Sz(1-z)(2z-1),\\
\phi_S^s(z,\xi,\omega^2)&=&\frac{1}{2\sqrt{2Nc}}F_S(\omega^2),\\
\phi_S^t(z,\xi,\omega^2)&=&\frac{1}{2\sqrt{2Nc}}F_S(\omega^2)(1-2z).
\end{eqnarray}
The dependence on $\xi$ does not show up in above functions, just because the Legendre polynomial $P_0(2\xi-1)$ is unity for the $S$ wave. The Gegenbauer moment $a_S$ is set to be $-0.8$, which is determined by the experimental data \cite{Lees:2011nf}. Note that we here only adopt the asymptotic form because the reliable theoretical studies are still absent. $F_S(\omega^2)$ is the $S$-wave time-like form factor containing the interaction between the two kaons in the $K\bar K$-pair. For most resonances, the form factors are usually taken to be relativistic Breit-Wigner (RBW) line shapes \cite{Tanabashi:2018oca}:
\begin{eqnarray}
F_S(\omega^2)=\frac{m_j^2}{m_j^2-\omega^2-im_j\Gamma_j(\omega)},
\label{RBW}
\end{eqnarray}
with the nominal mass $m_j$ being the mass of the resonance. $\Gamma(\omega)$ is the mass-dependent width.  In the general case of a spin-$L$ resonance, $\Gamma(\omega)$ can be expressed as
\begin{eqnarray}
\Gamma_j(\omega)=\Gamma_j^0\left(\frac{\mid\vec{q}\mid}{\mid\vec{q}_j\mid}\right)^{2L+1}
\left(\frac{m_j}{\omega}\right)X_L^2(\zeta),
\end{eqnarray}
where $\Gamma_j^0$ denotes the nominal width of the resonance. The value of $|\vec q|$ is the momentum of one of $K$ in the $K\bar K$-pair, which is valued $|\vec q_j|$ when $\omega=m_j$. The values of $\Gamma_j^0$ and $m_j$ can be found in ref.\cite{Tanabashi:2018oca}. $X_L(\zeta)$ is the Blatt-Weisskopf angular momentum barrier factor \cite{Blatt:1952ije}, whose expressions are given by
\begin{eqnarray}
&&L=0:\;\;X_L(\zeta)=1,\\
&&L=1:\;\;X_L(\zeta)=\sqrt{\frac{1+\zeta_0^2}{1+\zeta^2}},\\
&&L=2:\;\;X_L(\zeta)=\sqrt{\frac{9+3\zeta_0^2+\zeta_0^4}{9+3\zeta^2+\zeta^4}},
\end{eqnarray}
where $\zeta=r|\vec q|$ and $\zeta_0$ is the value of the $\zeta$ when the invariant mass of $K\bar K$-pair equals to the parent resonance. $L$ is the angular momentum of the kaon-pair, equaling to the spin of the corresponding resonance.  $r$ is the effective meson radius, which does not affect the results remarkably, so we take $r=4~\rm GeV^{-1}$ for all resonances.

In this work we shall consider the contributions from the scalar resonances $f_0(980)$, $f_0(1500)$ and $f_0(1710)$,  which are well analyzed by BaBar experiments \cite{Lees:2012kxa,Lees:2011nf}. The coefficients of the coherence summation of these three resonances in eq.(\ref{isobar}) are set to be $a_{f_0(980)}=2.9$, $a_{f_0(1500)}=1.0$, $a_{f_0(1710)}=0.5$, which have been determined by the experimental measurements\cite{Lees:2012kxa,Lees:2011nf}. Here, we suppose these coefficients are real, as we have not discussed the interferences among them.

For the $f_0(980)$, because there is an anomalous structure corresponding to the enhancement from the $KK$ system found around $980$ MeV in the $\pi^+\pi^-$ scattering \cite{AlstonGarnjost:1971kv, Flatte:1972rz}, it can be interpreted as a two-channel resonance combining the $\pi\pi$ and $KK$ channels. In the literatures, beside the Breit-Wigner (RBW) form, the Flatt\'{e} form \cite{Flatte:1976xu, Bugg:2008ig, Aaij:2014emv} is also usually applied to describe the line shape of $f_0(980)$, and it can be given as
\begin{eqnarray}\label{Flatte form}
F_S(\omega^2)=\frac{m_{f_0(980)}^2}{m_{f_0(980)}^2-\omega^2
-im_{f_0(980)}(g_{\pi\pi}\rho_{\pi\pi}+g_{KK}\rho_{KK}F_{KK}^2)},
\end{eqnarray}
where $g_{\pi\pi}$ and $g_{KK}$ are the $f_0(980)$ coupling constants to the $\pi\pi$ and $KK$ final states, respectively. The phase space factors $\rho_{\pi\pi}$ and $\rho_{KK}$ are parameterized as
\begin{eqnarray}
\rho_{\pi\pi}=\sqrt{1-\frac{4m_\pi^2}{\omega^2}},\,\,\,\,
\rho_{KK}=\sqrt{1-\frac{4m_K^2}{\omega^2}}.
\end{eqnarray}
The factor $F_{KK}=e^{-\alpha q^2}$ is to suppress the $K\overline K$ contribution with $\alpha \approx 2.0 ~\rm GeV^{-2} $ \cite{Aaij:2014emv}.

Next, we come to the $P$-wave two-kaon wave function. Because the third kaon in $B\to KKK$ decays is a pseudoscalar meson, so only the longitudinal polarization contribution is needed, and its form is very similar to the case of $\pi\pi$ pair and can be expressed as
\begin{eqnarray}
\Phi_P(KK)=\frac{1}{\sqrt{2N_c}}\left(p\mkern-8.5mu/\phi_P(z,\xi,\omega)+\omega\phi_P^s(z,\xi,\omega)
+\frac{p\mkern-8.5mu/_1p\mkern-8.5mu/_2-p\mkern-8.5mu/_2p\mkern-8.5mu/_1}{\omega(2\xi-1)}\phi_P^t(z,\xi,\omega)\right),
\end{eqnarray}
where $p$ is the momentum of the $K\bar K$-pair, while $p_{1(2)}$ is the momentum of one kaon in the $K\bar K$-pair. The corresponding twist-2 and 3 distribution amplitudes can be decomposed as the terms of Gegenbauer polynomials
\begin{eqnarray}
\phi_P^0(z,\xi,\omega)&=&\frac{3F_P^{\parallel}(\omega^2)}{\sqrt{2N_c}}z(1-z)\Big[1+a_P^0C_2^{3/2}(2z-1)\Big](2\xi-1),\\
\phi_P^s(z,\xi,\omega)&=&\frac{3F_P^{\perp}(\omega^2)}{2\sqrt{2N_c}}(1-2z)\Big[1+a_P^s(1-10z+10z^2)\Big](2\xi-1),\\
\phi_P^t(z,\xi,\omega)&=&\frac{3F_P^{\perp}(\omega^2)}{2\sqrt{2N_c}}(2z-1)^2\Big[1+a_P^tC_2^{3/2}(2z-1)\Big](2\xi-1),
\end{eqnarray}
with $a_P^0=-0.6$, $a_P^s=-0.8$, and $a_P^t=-0.3$. Also, $P$-wave time-like form factor $F_P^{\parallel(\perp)}$ describes the interaction between two kaons in $K\bar K$-pair. $F_P^{\parallel}$ can also taken to be the RBW line shape in eq. (\ref{RBW}), and $F_P^{\perp}$ can be obtained with the relation \cite{Wang:2016rlo}
 \begin{eqnarray}
  \frac{F_P^{\parallel}}{F_P^{\perp}}\approx \frac{f_V}{f_V^T},
 \label{perp}
 \end{eqnarray}
where $f_V$ and $f_V^T$ are the vector and tensor decay constants of the considered vector resonance. For the $B_{u,d}\to KKK$ decays, both Belle \cite{Garmash:2004wa} and BaBar \cite{Lees:2012kxa,Lees:2011nf} observed a narrow peak around $1.02 \rm GeV$ corresponding to the $\phi(1020)$ meson and measured the accurate branching fractions. As for the resonance $\phi(1680)$ meson, only the upper limit of the branching fraction of $B^+\to K^+ \phi(1680)\to K^+ K^+K^- $ decay has been reported by Belle\cite{Garmash:2004wa}. Since we have not enough data on it so far, we here only take the $\phi(1020)$ meson into account, the mass and width of which are referred to ref. \cite{Tanabashi:2018oca}.  For the decay constants of $\phi(1020)$, we take $f_{\phi(1020)}=(231\pm4)~\rm MeV$ and $f_{\phi(1020)}^T=(200\pm10)~ \rm MeV$, with scale $\mu=1.0~ \rm GeV$, the typical factorizable scale of $B$ decay.

At last, we will discuss the wave function of $D$-wave meson-pair in which the information of tensor meson resonances is included. As discussed in refs.\cite{Zou:2013wza, Zou:2012td, Zou:2012sx, Zou:2012sy, Cheng:2010yd, Wang:2010ni}, in $B$ meson decays involving a tensor in final states, the polarization components $\pm2$ of tensor meson do not contribute due to the conservation of the angular momentum. Therefore, for a tensor meson, a new introduced polarization vector $\epsilon^{\prime}$ associated with its the polarization tensor $\epsilon_{\mu\nu}$  makes its characters similar to the vector meson. Naturally, for $B\to KKK$ decays, the form of $D$-wave two-kaon wave function is the same as one of the $P$-wave, and can be decomposed as:
 \begin{eqnarray}
 \Phi_D(KK)=\frac{1}{\sqrt{2N_c}}\left(p\mkern-8.5mu/\phi_D(z,\xi,\omega)+\omega\phi_D^s(z,\xi,\omega)
+\frac{p\mkern-8.5mu/_1p\mkern-8.5mu/_2-p\mkern-8.5mu/_2p\mkern-8.5mu/_1}{\omega(2\xi-1)}\phi_D^t(z,\xi,\omega)\right).
\end{eqnarray}
The distribution amplitudes are given as
\begin{eqnarray}
\phi_D(z,\xi,\omega)
&=&\sqrt{\frac{2}{3}}\frac{9F_D^{\parallel}(\omega^2)}{\sqrt{2N_c}}z(1-z)a_D^0\Big[2z-1\Big]P_2(\xi),\\
\phi_D^s(z,\xi,\omega)
&=&-\sqrt{\frac{2}{3}}\frac{9F_D^{\perp}(\omega^2)}{4\sqrt{2N_c}}a_D^0\Big[1-6z+6z^2\Big]P_2(\xi),\\
\phi_D^t(z,\xi,\omega)
&=&\sqrt{\frac{2}{3}}\frac{9F_D^{\perp}(\omega^2)}{4\sqrt{2N_c}}a_D^0(2z-1)\Big[1-6z+6z^2\Big]P_2(\xi),
\end{eqnarray}
with $a_D^0=0.6$. The $\xi$ dependent space factor $P_2(\xi)$ can be written as
\begin{eqnarray}
P_2(\xi)=1-6\xi+6\xi^2.
\end{eqnarray}
$F_D^{\parallel}$ and $F_D^{\perp}$ are the $D$-wave time-like form factors. Similarly, we also describe the $F_D^{\parallel}$ using the RBW line shape as eq.(\ref{RBW}), and determine the $F_D^{\perp}$ by the similar relation as eq.(\ref{perp}). The decay constants of $f_2^{\prime}(1525)$ can be taken as $f_{f_2^{\prime}(1525)}=126~\rm MeV$ and $f^T_{f_2^{\prime}(1525)}=65~\rm MeV$. Since there are no sufficient experiment measurements and reliable theoretical studies on the decay constants of $f_2(2010)$, we then define a ratio as
\begin{eqnarray}
r_t=\frac{f^T_{f_2(2010)}}{f_{f_2(2010)}}
\end{eqnarray}
and left it as a free parameter. From the experimental results \cite{Lees:2012kxa, Lees:2011nf}, we can constrain it to be about $0.9\pm 0.1$.

\section{Perturbative Calculation}\label{sec:amplitude}
For simplicity, we work in the rest frame of the $B$ meson. In the light-cone coordinates, one can write the $B$ meson momentum $p_B$ and the light spectator quark momentum $k_B$ as
\begin{eqnarray}
p_B=\frac{m_B}{\sqrt2}(1,1,0_{\rm T}),
k_B=\left(\frac{m_B}{\sqrt2}x_1,0,k_{1{\rm T}}\right),
\end{eqnarray}
with $m_B$ being the $B$ meson mass and $x_1$ the momentum fraction. For the $B_{u,d}\to KR \to K(KK)$ decays, we define the resonant state momentum $p$ (in the plus $z$ direction), the associated spectator quark momentum $k$, the bachelor kaon momentum $p_3$ (in the minus $z$ direction) and the associated non-strange quark momentum $k_3$ as
\begin{eqnarray}\label{def-pp3}
p=\frac{m_B}{\sqrt2}(1,\eta^2,0_{\rm T}),&\quad&
k=\left(\frac{m_B}{\sqrt2}z,0,k_{\rm T}\right), \nonumber\\
p_3=\frac{m_B}{\sqrt2}(0,1-\eta^2,0_{\rm T}),&\quad&
k_3=\left(0,\frac{m_B}{\sqrt2}(1-\eta^2)x_3,k_{3{\rm T}}\right),
\end{eqnarray}
with the variable $\eta=w /m_B$, and the momentum fractions $z$ and $x_3$. So, the momenta $p_1$ and $p_2$ for the two kaons from the resonant state have the components
\begin{eqnarray}
p^+_1=\zeta\frac{m_B}{\sqrt2},           \quad    p^-_1=(1-\zeta)\eta^2\frac{m_B}{\sqrt2},
\quad p^+_2=(1-\zeta)\frac{m_B}{\sqrt2}, \quad    p^-_2=\zeta\eta^2\frac{m_B}{\sqrt2}.
\label{def-pp4}
\end{eqnarray}

\begin{figure}[!htb]
\begin{center}
\includegraphics[scale=1.25]{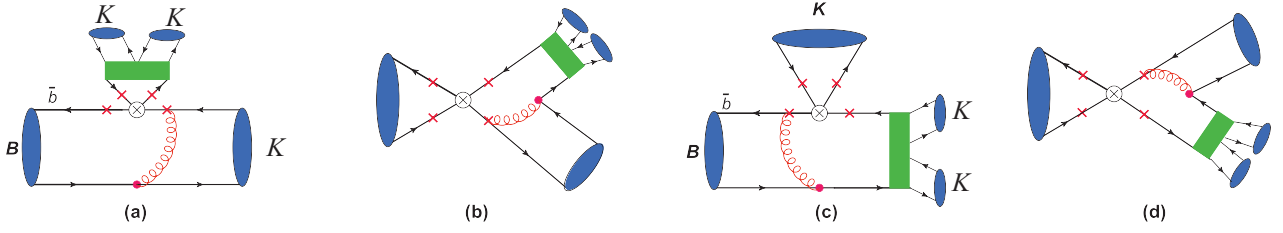}
\caption{Typical Feynman diagrams for the quasi-two-body decays $B\to KR\to KKK$ in PQCD,
in which the symbol $\otimes$ stands for the weak vertex, $\times$ denotes
possible attachments of hard gluons, and the green rectangle represents intermediate
states $R$.}\label{feynman}
\end{center}
\end{figure}

According to the effective Hamiltonian, we can draw the Feynman diagrams for the quasi-two-body decays $B\to KR\to KKK$ as shown in Fig.\ref{feynman}, where the symbol $\otimes$ stands for the weak vertex, $\times$ denotes possible attachments of hard gluons, and the green rectangle represents intermediate states $R$. In diagram (a) and (b), the spectator quark enters to the bachelor kaon, while it comes to the kaon-pair or the resonance in diagrams (c) and (d). Using the two-kaon wave function, in the PQCD framework we perform the perturbative calculation of the quasi-two-body $B_{u,d}\to KR \to K(KK)$ decays and get the analytic decay amplitudes for each diagram with different operators. In this work, we will not introduce the concept of PQCD in detail, and we refer the readers to refs.\cite{Keum:2000ph, Lu:2000em, Yu:2005rh, Ali:2007ff}.


In Figure.\ref{feynman}(a), when the hard gluon is emitted from the heavy quark or the new produced collinear quark, the decay amplitudes can be factorized as the convolution of the local form factors $F_{S,P,D}$ and $B\to K$ transition form factor. For the sake of brevity, we here take the $S$-wave as an example for illustration. For the $S$-wave resonance, due to the fact that the neutral scalar mesons can not be produced through the $V\pm A$ currents, there only exist amplitudes with $S\pm P$ currents for these two cases, and the total amplitudes can be written as
\begin{multline}
\mathcal{F}_{KK}^{SP}=16C_F\pi m^4_B\int_0^1 dx_1dx_3\int_0^{\infty}b_1db_1b_3db_3F_S \phi_{B}(x_1,b_1)\eta\\ \Bigg\{\Big[-\phi_K^a(x_3)
+x_3 r_K\phi_K^t(x_3)-(2+x_3)r_K\phi_K^p(x_3)\Big]E_{ef}(t_a)h_{ef}(x_1,x_3(1-\eta^2),b_1,b_3)\\
+\Big[2(\eta^2-1)r_K\phi_K^p(x_3)\Big]E_{ef}(t_b)h_{ef}(x_3,x_1(1-\eta^2),b_3,b_1)\Bigg\},
\end{multline}
where $r_K=m_{0K}/m_B$ with chiral mass of kaon $m_{0K}$. $b_i$ is the conjugate variable of the transverse momentum $k_{iT}$. $\phi_K^{a,p,t}$ are the distribution amplitudes of the kaon. The Sudakov form factor $E_{ef}$ and the hard function $h_{ef}$ can be found in ref.\cite{Zou:2015iwa}. When the gluon comes from two quarks of the bachelor kaon, that is the so-called nonfactorizable hard-scattering diagram, the amplitudes involve all the wave functions including the $B$, $K$, and kaon-pair wave functions and become complicated. If the $(V-A)(V-A)$ current is inserted, the total amplitude is written as
\begin{multline}
\mathcal{M}_{KK}^{LL}=16\sqrt{\frac{2}{3}}C_F\pi m^4_B\int_0^1 dx_1dzdx_3\int_0^{\infty}b_1db_1b_zdb_z\phi_B(x_1,b_1)\phi_S(z)\\
\Bigg\{\Big[(z-1)\phi_K^a(x_3)
+r_K\Big(x_3(\phi_K^t(x_3)-\phi_K^p(x_3))\\
+\eta^2\Big((z-x_3)\phi_K^t(x_3)+(z+x_3-2)\phi_K^p(x_3)\Big)\Big)\Big]
 E_{enf}(t_c)h_{enf}(\alpha,\beta_1,b_1,b_z) \\
-\Big[(z+x_3)\phi_K^a(x_3)-\eta^2(z+2x_3)\phi_K^a(x_3)-r_K\Big(x_3(\phi_K^p(x_3)+\phi_K^t(x_3)) \\
 -\eta^2\Big((x_3-z)\phi_K^p(x_3)+(x_3+z)\phi_K^t(x_3)\Big)\Big)\Big]E_{enf}(t_d)h_{enf}(\alpha,\beta_2,b_1,b_z)\Bigg\},
\end{multline}
where the related functions are also found in ref.\cite{Zou:2015iwa}. The amplitudes with $(V-A)(V+A)$ and $(S-P)(S+P)$ currents are also given respectively as
\begin{multline}
\mathcal{M}_{KK}^{LR}=16\sqrt{\frac{2}{3}}C_F\pi m_B^4\eta \int_0^1dx_1dzdx_3\int_0^{\infty}b_1db_1b_zdb_z\phi_B(x_1,b_1)\\
\Bigg\{\Big[(z-1)(\eta^2-1)\phi_K^a(x_3)\Big(\phi_S^s(z)+\phi_S^t(z)\Big)
+r_K\Big((1-z)(\phi_S^s(z)+\phi_S^t(z))(\phi_K^p(x_3)+\phi_K^t(x_3))\\
+(x_3+(1-x_3)\eta^2)(\phi_K^t(x_3)+\phi_K^p(x_3))(\phi_S^s(z)-\phi_S^t(z))\Big)\Big] E_{enf}(t_c)h_{enf}(\alpha,\beta_1,b_1,b_z)\\
+\Big[z(\eta^2-1)\phi_K^a(x_3)\Big(\phi_S^s(z)-\phi_S^t(z)\Big)
+r_K\Big(z(\phi_K^t(x_3)-\phi_K^p(x_3))(\phi_S^s(z)-\phi_S^t(z)) \\
+x_3(\eta^2-1)(\phi_K^p(x_3)+\phi_K^t(x_3))(\phi_S^s(z)+\phi_S^t(z))\Big)\Big]  E_{enf}(t_d)h_{enf}(\alpha,\beta_2,b_1,b_z)\Bigg\},
\end{multline}
\begin{multline}
\mathcal{M}_{KK}^{SP}=16\sqrt{\frac{2}{3}}C_F\pi m_B^4\int_0^1dx_1dzdx_3\int_0^{\infty}b_1db_1b_zdb_z \phi_B(x_1,b_1)\phi_S(z) \\
\Bigg\{\Big[(1-z+x_3)\phi_K^a(x_3)-r_Kx_3(\phi_K^p(x_3)+\phi_K^p(x_3)) +r_K\eta^2\Big((x_3-z)\phi_K^t(x_3) \\
+(z+x_3-2)\phi_K^p(x_3)\Big)\Big]E_{enf}(t_c)h_{enf}(\alpha,\beta_1,b_1,b_z)  \\
-\Big [z\phi_K^a(x_3)+r_Kx_3(\phi_K^t(x_3)-\phi_K^p(x_3))-r_K\eta^2\Big((z-x_3)\phi_K^p(x_3)\\
  +(z+x_3)\phi_K^t(x_3)\Big) \Big]E_{enf}(t_d)h_{enf}(\alpha,\beta_2,b_1,b_z)\Bigg \}.
\end{multline}
Note that in the charmless $B\to PP$ decays with $P$ denoting a pseudoscalar meson, the contributions from the nonfactorizable hard-scattering diagrams are always highly cancelled by each other, because of the negative relative sign caused by two quark propagators. So, in that case, these contributions are suppressed. However, in the current cases, because the distribution amplitudes of meson-pair are antisymmetric, the contributions are not suppressed but enhanced and provide remarkable contributions.

In Figure.\ref{feynman}(b), it is called the annihilation diagram. In term of the attachments of the hard gluon, the diagrams can be similarly classed into two kinds, the factorizable annihilation diagrams and the nonfactorizable annihilation ones, namely. For the factorizable ones, when we insert the $(V-A)(V-A)$, $(V-A)(V+A)$ and $(S-P)(S+P)$ currents, we then obtain the amplitudes as
\begin{multline}
\mathcal{A}_{KK}^{LL}=-8C_F\pi f_B m_B^4\int_0^1dzdx_3\int_0^{\infty}b_zdb_zb_3db_3\Bigg\{\Big[(1-\eta^2+x_3(2\eta^2-1))\phi_K^a(x_3)\phi_S(z) \\
 +2 r_K\eta\Big(x_3\phi_K^t(x_3)+(2-x_3)\phi_K^p(x_3)\Big)\phi_S^s(z)\Big]E_{af}(t_e)h_{af}(\alpha_1,\beta,b_z,b_3) \\
 +\Big[z(\eta^2-1)\phi_K^a(x_3)\phi_S(z)-2 r_K\eta\phi_K^p(x_3)\Big((1+z)\phi_S^s(z) \\-(1-z)\phi_S^t(z)\Big)\Big]
 E_{af}(t_f)h_{af}(\alpha_2,\beta,b_z,b_3)\Bigg\},
\end{multline}
\begin{eqnarray}
\mathcal{A}_{KK}^{LR}=-\mathcal{A}_{KK}^{LL},
\end{eqnarray}
\begin{multline}
\mathcal{A}_{KK}^{SP}=16C_F\pi m_B^4f_B\int_0^1dzdx_3\int_0^{\infty}b_zdb_zb_3db_3
\Bigg\{\Big[2\eta\phi_K^a(x_3)\phi_S^s(z) \\
 +r_K\Big((x_3-1)(\eta^2-1)\phi_K^t(x_3)+(1+\eta^2+x_3(\eta^2-1))\phi_K^p(x_3)\Big)\phi_S(z)\Big]  E_{af}(t_e)h_{af}(\alpha_1,\beta,b_z,b_3) \\
 +\Big[2(1+\eta^2(z-1))r_K\phi_K^p(x_3)\phi_S(z)+z\eta\phi_K^a(x_3)\Big(\phi_S^s(z)-\phi_S^t(z)\Big)\Big]   E_{af}(t_f)h_{af}(\alpha_2,\beta,b_z,b_3)\Bigg\}.
\end{multline}
As for the nonfactorizable annihilation diagrams, the amplitudes with different currents are calculated as
\begin{multline}
\mathcal{W}_{KK}^{LL}=16\sqrt{\frac{2}{3}}C_F\pi m_B^4\int_0^1dx_1dzdx_3 \int_0^{\infty}b_1db_1b_zdb_z\phi_B(x_1,b_1) \Bigg \{\Big[-z\phi_K^a(x_3)\phi_S(z) \\
 +r_K\eta\Big(\phi_K^t(x_3)(\phi_S^t(z)(z-x_3-1)+\phi_S^s(z)(z+x_3-1)) \\
 +\phi_K^p(x_3)(\phi_S^s(z)(x_3-z-3)+\phi_S^t(1-z-x_3))\Big)\Big]
 E_{anf}(t_g)h_{anf}(\alpha,\beta_1,b_1,b_z) \\
 +\Big[(1-x_3)\phi_K^a(x_3)\phi_S(z)+r_K\eta\Big((1-x_3)(\phi_K^p(x_3)-\phi_K^t(x_3))(\phi_S^s(z)+\phi_S^t(z))\\
 +z(\phi_K^p(x_3)+\phi_K^t(x_3))(\phi_S^s(z)-\phi_S^t(z))\Big)
 \Big]E_{anf}(t_h)h_{anf}(\alpha,\beta_2,b_1,b_z)\Bigg \},
\end{multline}
\begin{multline}
\mathcal{W}_{KK}^{LR}=16\sqrt{\frac{2}{3}}C_F\pi m_B^4\int_0^1dx_1dzdx_3\int_0^{\infty}b_1db_1b_zdb_z\phi_B(x_1,b_1)\\
 \Bigg\{\Big[(2-z)\phi_K^a(x_3)(\phi_S^s(z)+\phi_S^t(z))+r_K\phi_S(z)\Big(\phi_K^p(x_3)[-1+x_3(\eta^2-1) +\eta^2(z-3)] \\
+\phi_K^t(x_3)[(1+x_3)(1-\eta^2)+\eta^2z]\Big)\Big]E_{anf}(t_g)h_{anf}(\alpha,\beta_1,b_1,b_z) \\
 +\Big[z\eta\phi_K^a(x_3)(\phi_S^s(z)+\phi_S^t(z))+r_K\phi_S(z)\Big(\phi_K^t(x_3)[1-x_3-(1+z-x_3)\eta^2]\\
 -\phi_K^p(x_3)[1-x_3 +(x_3+z-1)\eta^2]\Big)\Big]E_{anf}(t_h)h_{anf}(\alpha,\beta_2,b_1,b_z)\Bigg\},
\end{multline}
\begin{multline}
\mathcal{W}_{KK}^{SP}=16\sqrt{\frac{2}{3}}C_F\pi m_B^4\int_0^1dx_1dzdx_3\int_0^{\infty}b_1db_1b_zdb_z\phi_B(x_1,b_1)\\
\Bigg \{\Big[(1-x_3)\phi_K^a(x_3)\phi_S(z)+r_K\eta\Big(\phi_K^t(x_3)[\phi_S^t(z)(1+x_3-z)+\phi_S^s(z)(z+x_3-1)] \\
 +\phi_K^p(x_3)[\phi_S^t(z)(1-x_3-z)+\phi_S^s(z)(3-x_3+z)]\Big)\Big]E_{anf}(t_g)h_{anf}(\alpha,\beta_1,b_1,b_z) \\
 -\Big[z\phi_K^a(x_3)\phi_S(z)+r_K\eta\Big((1-z)(\phi_K^p(x_3)+\phi_K^t(x_3)(\phi_S^s(z)-\phi_S^t(z)) \\
 +z(\phi_K^p(x_3)-\phi_K^t(x_3))(\phi_S^s(z)+\phi_S^t(z)))\Big)\Big]E_{anf}(t_h)h_{anf}(\alpha,\beta_2,b_1,b_z)\Bigg\}
\end{multline}

In Figure.\ref{feynman}(c), the bachelor $K$ meson is emitted and the spectator quark flows into the kaon-pair. Accordingly, we have the factorizable and nonfactorizable contributions. For the factorizable diagrams, the amplitudes can be factorized as the convolution of the kaon decay constant and the $B\to KK$ transition factor. With different currents $(V-A)(V-A)$ and $(S-P)(S+P)$, the whole amplitudes can be read as
\begin{multline}
\mathcal{F}_K^{LL}=8C_Ff_K\pi m_B^4\int_0^1dx_1dz\int_0^{\infty}b_1db_1b_zdb_z\phi_B(x_1,b_1)(1-\eta^2) \\
 \Bigg\{ \Big[(1+z)\phi_S(z)-(2z-1)\eta\Big(\phi_S^s(z)+\phi_S^t(z)\Big) \Big]E_{ef}(t_a)h_{ef}(x_1,z,b_1,b_z), \\
 + \Big[ 2\eta\phi_S^s(z)+\eta^2\phi_S(z) \Big]E_{ef}(t_b)h_{ef}(z,x_1,b_z,b_1)\Bigg\},
\end{multline}
\begin{multline}
\mathcal{F}_K^{SP}=16C_Ff_K\pi r_K m^4_B\int_0^1dx_1dz\int_0^{\infty}b_1db_1b_zdb_z\phi_B(x_1,b_1) \\
 \Bigg\{ \Big[z\phi_S^t(z)-(2+z)\eta\phi_S^s(z)-(1+(1-2z)\eta^2)\phi_S(z) \Big]E_{ef}(t_a)h_{ef}(x_1,z,b_1,b_z), \\
 -\Big[2\eta\phi_S^s(z)-2\eta^2 \phi_S(z)\Big]E_{ef}(t_b)h_{ef}(z,x_1,b_z,b_1)\Bigg\}.
\end{multline}
Because the $(V-A)(V+A)$ current has no effect on the decay concerned, we will not list its amplitude here. For the nonfactorizable diagrams, the hard gluon comes from one of the two quarks of the bachelor kaon, and then kick the spectator. In this case, the amplitudes $\mathcal{M}_K^{LL,LR,SP}$ with different currents are listed as
\begin{multline}
\mathcal{M}_K^{LL}=16\sqrt{\frac{2}{3}}C_F\pi m_B^4\int_0^1dx_1dx_3dz\int_0^{\infty}b_1db_1b_3db_3 \phi_B(x_1,b_1)\phi_K^a(x_3) \\
\Bigg \{\Big[z\eta(\phi_S^t(z)-\phi_S^s(z)) +(1-x_3+(z+2x_3-2)\eta^2)\phi_S(z)\Big]E_{enf}(t_c)h_{enf}(\alpha,\beta_1,b_1,b_3), \\
+\Big[z\eta(\phi_S^s(z)+\phi_S^t(z))-(z+x_3-(z+x_3)\eta^2)\phi_S(z)\Big]
E_{enf}(t_d)h_{enf}(\alpha,\beta_2,b_1,b_3)\Bigg\},
\end{multline}
\begin{multline}
\mathcal{M}_K^{LR}=-16\sqrt{\frac{2}{3}}C_F\pi r_K m_B^4\int_0^1dx_1dx_3dz
\int_0^{\infty}b_1db_1b_3db_3\phi_B(x_1,b_1) \\
\Bigg\{\Big[\phi_K^t(x_3)\Big(\phi_S(z)(1-x_3+(x_1-z-1)\eta^2)-\eta(\phi_S^t(z)(1+z-x_3) \\
 +\phi_S^s(z)(x_3+z-1))\Big)+\phi_K^p(x_3)\Big(\phi_S(z)(1-x_3+(x_3+z-1)\eta^2) \\
 +\eta(\phi_S^s(z)(1+z-x_3)+\phi_S^t(z)(x_3+z-1))\Big)\Big]E_{enf}(t_c)h_{enf}(\alpha,\beta_1,b_1,b_3) \\
 +\Big[\phi_S(\phi_K^t(x_3)-\phi_K^p(x_3))x_3-((\phi_K^p(x_3)+\phi_K^t(x_3))(\phi_S^s(z)+\phi_S^t(z))z \\
 +(\phi_K^p(x_3)-\phi_K^t(x_3))(\phi_S^s(z)-\phi_S^t(z))x_3)\eta-\phi_S(z)(\phi_K^p(x_3)(z-x_3) \\
 -\phi_K^t(x_3)(z+x_3))\eta^2\Big]E_{enf}(t_d)h_{enf}(\alpha,\beta_2,b_1,b_3)\Bigg\},
\end{multline}
\begin{multline}
\mathcal{M}_K^{SP}=16\sqrt{\frac{2}{3}}C_F\pi m_B^4 \int_0^1dx_1dx_3dz\int_0^{\infty}b_1db_1b_3db_3\phi_B(x_1,b_1)\phi_K^a(x_3) \\
\Bigg \{\Big[z\eta\Big(\phi_S^s(z)+\phi_S^t(z)\Big)
+\phi_S(z)(x_3-z-1+(2+z-2x_3)\eta^2)\Big]E_{enf}(t_c)h_{enf}(\alpha,\beta_1,b_1,b_3) \\
 -\Big[z\eta \Big(\phi_S^s(z)-\phi_S^t(z)\Big)-\phi_S(z)(x_3+(z-2x_3)\eta^2)
\Big]E_{enf}(t_d)h_{enf}(\alpha,\beta_2,b_1,b_3)\Bigg\}.
\end{multline}

Similar to Figure.\ref{feynman}(b), we can draw another annihilation diagram as shown in diagram Figure.\ref{feynman}(d). Then, we can calculate the related amplitudes of factorizable and nonfactorizable diagrams with possible currents as
\begin{multline}
\mathcal{A}_K^{LL}=-8C_F\pi f_Bm_B^4\int_0^1dx_3dz\int_0^{\infty}b_3db_3b_zdb_z
\Bigg\{\Big[(z-1)(\eta^2-1)\phi_K^a(x_3)\phi_S(z) \\
 +2\eta r_K\phi_K^p(x_3)\Big((z-2)\phi_S^s(z)-z\phi_S^t(z)\Big)\Big]E_{af}(t_e)h_{af}(\alpha_1,\beta,b_3,b_z) \\
 +\Big[\Big(-x_3+(2x_3-1)\eta^2\Big)\phi_K^a(x_3)\phi_S(z)+2\eta r_K\phi_S^s(z)\Big((x_3-1)\phi_K^t(x_3) \\
 +(x_3+1)\phi_K^p(x_3)\Big)\Big]E_{af}(t_f)h_{af}(\alpha_2,\beta,b_3,b_z)\Bigg\},
\end{multline}
\begin{eqnarray}
\mathcal{A}_K^{LR}=-\mathcal{A}_K^{LL},
\end{eqnarray}
\begin{multline}
\mathcal{A}_K^{SP}=16C_Ff_B\pi m_B^4\int_0^1dx_3dz\int_0^{\infty}b_3db_3b_zdb_z
\Bigg\{\Big[\phi_K^a(x_3)(\phi_S^s(z)+\phi_S^t(z))(1-z)\eta \\
 -2r_K\phi_K^p(x_3)\phi_S(z)(1-(z-1)\eta^2)\Big]E_{af}(t_e)h_{af}(\alpha_1,\beta,b_3,b_z) \\
 +\Big[2\phi_K^a(x_3)\phi_S^s(z)\eta-r_K\phi_S(z)(\phi_K^t(x_3)x_3(\eta^2-1) \\
 +\phi_K^p(x_3)(2\eta^2+x_3(1-\eta^2)))\Big]E_{af}(t_f)h_{af}(\alpha_2,\beta,b_3,b_z)\Bigg\}.
\end{multline}
\begin{multline}
\mathcal{W}_K^{LL}=16\sqrt{\frac{2}{3}}\int_0^1dx_1dx_3dz\int_0^{\infty}b_1db_1b_3db_3\phi_B(x_1,b_1)
\Bigg\{\Big[\phi_K^a(x_3)\phi_S(z)(-x_3+(2x_3+z-1)\eta^2) \\
 +r_K\eta\Big(\phi_K^t(x_3)(\phi_S^t(z)(1+z-x_3)+\phi_S^s(z)(z+x_3-1))+\phi_K^p(x_3)(\phi_S^s(z)(3-z+x_3) \\
 +\phi_S^t(z)(1-x_3-z))\Big)\Big]E_{anf}(t_g)h_{anf}(\alpha,\beta_1,b_1,b_3) \\
 +\Big[\phi_K^a(x_3)\phi_S(z)(1-z)-r_K\eta\Big((\phi_K^p(x_3)+\phi_K^t(x_3))(\phi_S^t(z)-\phi_S^s(z))(z-1) \\
 +(\phi_K^p(x_3)-\phi_K^t(x_3))(\phi_S^s(x_3)+\phi_S^t(x_3))x_3\Big)\Big]
 E_{anf}(t_h)h_{anf}(\alpha,\beta_2,b_1,b_3)\Bigg\}.
\end{multline}
\begin{multline}
\mathcal{W}_K^{LR}=16\sqrt{\frac{2}{3}}\int_0^1dx_1dx_3dz\int_0^{\infty}b_1db_1b_3db_3\phi_B(x_1,b_1)\\
\Bigg\{\Big[(1+z)\eta\phi_K^a(x_3)(\phi_S^t(z)-\phi_S^s(z))
 +\phi_S(z)r_K\Big(\phi_K^p(x_3)(x_3-2-(x_3+z)\eta^2) \\ +\phi_K^t(x_3)(x_3-2+(2+z-x_3)\eta^2)\Big)\Big]E_{anf}(t_g)h_{anf}(\alpha,\beta_1,b_1,b_3) \\
 +\Big[(z-1)\eta\phi_K^a(x_3)(\phi_S^s(z)-\phi_S^t(z))-r_K\phi_S(z)\Big(\phi_K^p(x_3)(x_3-(x_3+z-2)\eta^2) \\
 +\phi_K^t(x_3)(x_3-(x_3-z)\eta^2)\Big)\Big]E_{anf}(t_h)h_{anf}(\alpha,\beta_2,b_1,b_3)\Bigg\},
\end{multline}
\begin{multline}
\mathcal{W}_K^{SP}=16\sqrt{\frac{2}{3}}C_F\pi m_B^4 \int_0^1dx_1dx_3dz \int_0^{\infty}b_1db_1b_3db_3 \phi_B(x_1,b_1)\\
\Bigg\{\Big[(1-z-\eta^2)\phi_K^a(x_3)\phi_S(z)-r_K\eta\Big(\phi_K^t(x_3)
(\phi_S^s(z)(1-z-x_3)+\phi_S^t(z)(1+z-x_3)) \\
 +\phi_P^p(x_3)(\phi_S^s(x_3)(3-z+x_3)+\phi_S^t(z)(x_3+z-1))\Big)\Big]
E_{anf}(t_g)h_{anf}(\alpha,\beta_1,b_1,b_3) \\
 +\Big[\Big(-x_3+(2x_3+z-2)\eta^2\Big)\phi_K^a(x_3)\phi_S(z)+r_K\eta\Big(\phi_K^t(x_3)(\phi_S^s(z)(z-1+x_3) \\
 +\phi_S^t(z)(z-1-x_3))+\phi_K^p(x_3)(\phi_S^s(z)(1-z+x_3) \\+\phi_S^t(z)(1-z-x_3))\Big)\Big] E_{anf}(t_h)h_{anf}(\alpha,\beta_2,b_1,b_3)\Bigg\}.
\end{multline}

For the $S$-wave resonance $f_0(980)$, the inner quark structure is very complicated. Though many data showed that it may be four-quark state, we here regard it as the mixing state between two-quark states $q\bar q=(u\bar u+d\bar d)/\sqrt{2}$ and $s\bar s$ with mixing angle $\theta=40^\circ$. More details will be discussed in the following section. So, we can write down the total amplitudes of $B \to K f_0(980) \to K K^+K^-$ with the Wilson coefficients and the CKM matrix elements as
 \begin{eqnarray}
 {\cal A}(B^0 \to K^0 f_0(980) \to K^0 K^+K^-)&=&{\cal M}_S^{n}[f_0(q\bar{q})]\sin\theta+{\cal M}_S^{n}[f_0(s\bar{s})]\cos\theta,\label{amp9801}\\
 {\cal A}(B^+ \to K^+ f_0(980) \to K^+ K^+K^-)&=&{\cal M}_S^{p}[f_0(q\bar{q})]\sin\theta+{\cal M}_S^{p}[f_0(s\bar{s})]\cos\theta,\label{amp9802}
 \end{eqnarray}
where the expressions of ${\cal M}^{n,p}[f_0(q\bar{q})])$ and ${\cal M}^{n,p}[f_0(s\bar{s})]$ are
 \begin{multline}
 {\cal M}_S^{n}[f_0(q\bar{q})]=\frac{G_F}{2}\Bigg\{V_{ub}^*V_{us}C_2\mathcal{M}_{KK}^{LL}-V_{tb}^*V_{ts}\Big[
 \left(2C_4+\frac{1}{2}C_{10}\right)\mathcal{M}_{KK}^{LL} \\
 +\left(2C_6+\frac{1}{2}C_8\right)\mathcal{M}_{KK}^{SP}
   +\left(\frac{1}{3}C_3+C_4-\frac{1}{6}C_9-\frac{1}{2}C_{10}\right)
   \left(\mathcal{F}_K^{LL}+\mathcal{A}_K^{LL}\right) \\
 +\left(\frac{1}{3}C_5+C_6-\frac{1}{6}C_7-\frac{1}{2}C_8\right)
  \left(\mathcal{F}_K^{SP}+\mathcal{A}_K^{SP}\right) \\
  +\left(C_3-\frac{1}{2}C_9\right)\left(\mathcal{M}_{K}^{LL}+\mathcal{W}_K^{LL}\right)
   +\left(C_5-\frac{1}{2}C_7\right)\left(\mathcal{M}_K^{LR}+\mathcal{W}_K^{LR}\right)\Big]\Bigg\},
 \end{multline}
 \begin{multline}
 {\cal M}_S^{n}[f_0(s\bar{s})]=-\frac{G_F}{\sqrt{2}}V_{tb}^*V_{ts}\Big[
 \left(\frac{1}{3}C_5+C_6-\frac{1}{6}C_7-\frac{1}{2}C_8\right)\mathcal{M}_{KK}^{LR}
 +\left(C_5-\frac{1}{2}C_7\right)\mathcal{F}_{KK}^{SP} \\
  +\left(C_3+C_4-\frac{1}{2}C_9-\frac{1}{2}C_{10}\right)\mathcal{M}_{KK}^{LL}
   +\left(C_6-\frac{1}{2}C_8\right)\mathcal{M}_{KK}^{SP} \\
 +\left(\frac{1}{3}C_3+C_4-\frac{1}{6}C_9-\frac{1}{2}C_{10}\right)\mathcal{A}_{KK}^{LL}
   +\left(\frac{1}{3}C_5+C_6-\frac{1}{6}C_7-\frac{1}{2}C_8\right)\mathcal{A}_{KK}^{SP} \\
 +\left(C_3-\frac{1}{2}C_9\right)\mathcal{W}_{KK}^{LL}
   +\left(C_5-\frac{1}{2}C_7\right)\mathcal{W}_{KK}^{LR}\Big],
 \end{multline}
\begin{multline}
{\cal M}_S^{p}[f_0(q\bar{q})]=\frac{G_F}{2}\Bigg\{V_{ub}^*V_{us}\Big[C_2\mathcal{M}_{KK}^{LL}
+\left(\frac{1}{3}C_1+C_2\right)\left(\mathcal{F}_K^{LL}+\mathcal{A}_K^{LL}\right) \\
+C_1\left(\mathcal{M}_K^{LL}+\mathcal{W}_K^{LL}\right)\Big]
-V_{tb}^*V_{ts}\Big[\left(2C_4+\frac{1}{2}C_{10}\right)\mathcal{M}_{KK}^{LL}
+\left(2C_6-\frac{1}{2}C_8\right)\mathcal{M}_{KK}^{SP} \\
+\left(\frac{1}{3}C_3+C_4
+\frac{1}{3}C_9+C_{10}\right)\left(\mathcal{F}_K^{LL}+\mathcal{A}_K^{LL}\right)
+\left(C_3+C_9\right)\left(\mathcal{M}_K^{LL}+\mathcal{W}_K^{LL}\right) \\
+\left(\frac{1}{3}C_5+C_6+\frac{1}{3}C_7+C_8\right)\left(\mathcal{F}_K^{SP}
+\mathcal{A}_K^{SP}\right)+\left(C_5+C_7\right )\left(\mathcal{M}_K^{LR}+\mathcal{W}_K^{LR}\right)\Big]\Bigg\},
\end{multline}
\begin{multline}
{\cal M}_S^{p}[f_0(s\bar{s})]=\frac{G_F}{\sqrt{2}}\Bigg\{V_{ub}^*V_{us}\Big[
\left(\frac{1}{3}C_1+C_2\right)\mathcal{A}_{KK}^{LL}+C_2\mathcal{W}_{KK}^{LL}\Big] \\
-V_{tb}^*V_{ts}\Big[\left(\frac{1}{3}C_5+C_6-\frac{1}{6}C_7-\frac{1}{2}C_8\right)\mathcal{F}_{KK}^{SP}
+\left(C_3+C_4-\frac{1}{2}C_9-\frac{1}{2}C_{10}\right)\mathcal{M}_{KK}^{LL} \\
+\left(C_5-\frac{1}{2}C_7\right)\mathcal{M}_{KK}^{LR}+\left(C_6-\frac{1}{2}C_8\right)\mathcal{M}_{KK}^{SP}
+\left(\frac{1}{3}C_3+C_4+\frac{1}{3}C_9+C_{10}\right)\mathcal{A}_{KK}^{LL} \\
+\left(\frac{1}{3}C_5+C_6+\frac{1}{3}C_7+C_8\right)\mathcal{A}_{KK}^{SP}+\Big(C_3+C_9\Big)\mathcal{W}_{KK}^{LL}
+\Big(C_5+C_7\Big)\mathcal{W}_{KK}^{LR}\Big]\Bigg\}.
\end{multline}

It should be emphasized that there are two positive kaon $B^+ \to K^+ K^+K^-$, but one of them is in the kaon-pair and the other is a bachelor in the quasi-two-body decay region. Once tracking the kaon with negative charge, these two  positive ones could be distinguishable in the experiments. With the total amplitude $\cal A$ and its conjugate $\overline{\cal A}$, we then give the definition of the direct $CP$ asymmetry as
 \begin{eqnarray}
 {\cal A}_{CP}=\frac{\overline{{\cal A}}-{\cal A}}{\overline{{\cal A}}+{\cal A}}.
 \end{eqnarray}

Similarly, we adopt the mixing forms discussed in ref.\cite{Cheng:2006hu} and write the total $B \to K f_0(1500) \to K K^+K^-$ and $B \to K f_0(1710) \to K K^+K^-$as
 \begin{eqnarray}
 {\cal A}(B^{0,+} \to K^{0,+} f_0(1500) \to K^{0,+} K^+K^-)&=&{\cal M}_S^{n,p}[f_0(q\bar{q})](-0.54)+{\cal M}_S^{n,p}[f_0(s\bar{s})](+0.84),\label{amp1500}\\
 {\cal A}(B^{0,+} \to K^{0,+} f_0(1710) \to K^{0,+} K^+K^-)&=&{\cal M}_S^{n,p}[f_0(q\bar{q})](+0.32)+{\cal M}_S^{n,p}[f_0(s\bar{s})](+0.18).\label{amp1710}
 \end{eqnarray}
Adopting the same strategy, we could calculate the total amplitudes of decays $B \to K K^+K^-$ with resonances $\phi(1020)$, $f_2^\prime (1525)$ and $f_2 (2010)$. Due to the space limited, we here do not present them any more.

At last, we write down the differential branching ratio for the quasi-two-body decay $B \to K  K^+K^-$ as,
\begin{eqnarray}
\frac{d{\mathcal B}}{dw^2}=\tau_B\frac{|\vec{p}_1||\vec{p}_3|}
{32\pi^3m^3_B}|{\mathcal A}|^2\;,
\label{eqn-bf}
\end{eqnarray}
$\tau_B$ being the $B$ meson mean lifetime. In the center-of-mass frame of the kaon pair, $|\vec{p}_1|$ and $|\vec{p}_3|$ are written as
\begin{eqnarray}
 |\vec{p}_1|=\frac{\sqrt{\lambda(\omega^2,m_{K}^2,m_{K}^2)}}{2 \omega}\;, \quad~~
 |\vec{p}_3|=\frac{\sqrt{\lambda(M^2,m_{K}^2,\omega^2)}}{2 \omega}\;,
\end{eqnarray}
with the kaon mass $m_K$ and the K\"{a}ll\'en function $\lambda(a,b,c)=a^2+b^2+c^2-2ab-2bc-2ac$.

\section{Numerical Results and Discussions}\label{sec:result}
In this section, let us first list the parameters used in our numerical calculations, such as the masses, lifetimes, and decay constants of the $B$ mesons, the CKM matrix elements and the QCD scale, and they are given as follows  \cite{Tanabashi:2018oca}:
\begin{eqnarray}
&&
m_{B}=5.279~ {\rm GeV},\;\;f_B=0.19\pm 0.02 ~{\rm GeV},\mid V_{tb}\mid=1.0,\mid V_{ts}\mid=0.04133\pm0.00074,\;\;\nonumber\\
&&\mid V_{ub}\mid=0.00365\pm0.00012,\;\;\mid V_{us}\mid=0.22452\pm0.00044,\;\;\nonumber\\
&& \;\;\tau_{B_u}/\tau_{B_d}=1.638/1.525 ~{\rm ps},\Lambda_{QCD}^{f=4}=0.25\pm0.05~{\rm GeV}.
\end{eqnarray}

\begin{table}[!t]
\caption{$CP$ averaged branching ratios (in $10^{-6}$) of $B\to K^{+/0} (R \to)  K^+ K^-/K_S K_S$ decays in PQCD approach together
with experimental data\cite{Lees:2012kxa,Lees:2011nf}. The results from a model based on the factorization approach(MFA) are from ref.\cite{Cheng:2013dua}.}
 \label{br}
\begin{center}
\begin{tabular}{l c c c}
 \hline \hline
 \multicolumn{1}{c}{Decay Modes}&\multicolumn{1}{c}{PQCD } &\multicolumn{1}{c}{EXP\cite{Lees:2012kxa,Lees:2011nf}}  &\multicolumn{1}{c}{MFA\cite{Cheng:2013dua}} \\
\hline\hline
 $B^+ \to K^+(\phi(1020)\to)K^+ K^-$
 &$3.81^{+1.44+0.64+0.27}_{-1.03-0.33-0.00}$
 &$4.48\pm0.22^{+0.33}_{-0.24}$
 &$2.9_{-0.0-0.5-0.0}^{+0.0+0.5+0.0}$\\

 $B^+ \to K^+(f_0(980)\to)K^+ K^-$
 &$10.13^{+5.60+2.22+0.71}_{-4.38-2.44-0.00}$
 &$9.4\pm1.6\pm2.8$
 &$11.0^{+0.0+2.6+0.0}_{-0.0-2.1-0.0}$\\

 $B^+ \to K^+(f_0(1500)\to)K^+ K^-$
 &$0.60^{+0.24+0.07+0.05}_{-0.24-0.06-0.02}$
 &$0.74\pm0.18\pm0.52$
 &$0.62^{+0.0+0.11+0.0}_{-0.0-0.10-0.0}$\\

 $B^+ \to K^+(f_0(1710)\to)K^+ K^-$
 &$1.64^{+0.89+0.42+0.08}_{-0.70-0.46-0.02}$
 &$1.12\pm0.25\pm0.50$
 &$1.1^{+0.0+0.2+0.0}_{-0.0-0.2-0.0}$\\

 $B^+ \to K^+(f_2^{\prime}(1525)\to)K^+ K^-$
 &$0.68^{+0.37+0.13+0.07}_{-0.29-0.14-0.00}$
 &$0.69\pm0.16\pm0.13$
 &$$\\

 $B^+ \to K^+(f_2(2010)\to)K^+ K^-$
 &$1.18^{+0.65+0.26+0.12}_{-0.50-0.19-0.00}$
 &$$
 &$$\\
 \hline
 $B^+ \to K^+(f_0(980)\to)K_S K_S$
 &$10.33^{+5.60+2.23+0.72}_{-4.38-2.44+0.00}$
 &$14.7\pm2.8\pm1.8$
 &$8.7^{+0.0+2.1+0.0}_{-0.0-1.6-0.0}$\\

 $B^+ \to K^+ (f_0(1500)\to)K_S K_S$
 &$0.59^{+0.24+0.07+0.05}_{-0.24-0.06-0.02}$
 &$0.42\pm0.22\pm0.58$
 &$0.59^{+0.00+0.10+0.00}_{-0.00-0.09-0.00}$ \\

 $B^+ \to K^+(f_0(1710)\to)K_S K_S$
 &$1.60^{+0.88+0.42+0.11}_{-0.70-0.45-0.01}$
 &$0.48^{+0.40}_{-0.24}\pm0.11$
 &$1.08^{+0.00+0.18+0.00}_{-0.00-0.17-0.00}$\\

 $B^+ \to K^+(f_2^{\prime}(1525)\to)K_S K_S$
 &$0.68^{+0.37+0.13+0.07}_{-0.29-0.13-0.00}$
 &$0.61\pm0.21^{+0.12}_{-0.09}$
 &$$\\

 $B^+ \to K^+(f_2(2010)\to)K_S K_S$
 &$0.69^{+0.36+0.14+0.07}_{-0.28-0.08-0.00}$
 &$$
 &$$\\

 \hline
  $B^0 \to K^0(\phi(1020)\to)K^+ K^-$
  &$3.22^{+1.36+0.48+0.18}_{-0.98-0.18-0.08}$
  &$3.48\pm0.28^{+0.21}_{-0.14}$
  &$2.6_{-0.0-0.4-0.0}^{+0.0+0.4+0.0}$\\

  $B^0 \to K^0(f_0(980)\to)K^+ K^-$
  &$9.10^{+5.12+2.19+0.69}_{-3.89-2.11-0.00}$
  &$7.0^{+2.6}_{-1.8}\pm2.4$
  &$9.1_{-0.0-1.4-0.0}^{+0.0+1.7+0.0}$\\

  $B^0 \to K^0(f_0(1500)\to)K^+ K^-$
  &$0.57^{+0.26+0.09+0.04}_{-0.22-0.15-0.00}$
  &$0.57^{+0.25}_{-0.19}\pm0.12$
  &$0.55_{-0.0-0.09-0.0}^{+0.0+0.10+0.0}$\\

  $B^0 \to K^0(f_0(1710)\to)K^+ K^-$
  &$1.48^{+0.82+0.39+0.11}_{-0.63-0.42-0.00}$
  &$4.4\pm0.7\pm0.5$
  &$1.0_{-0.0-0.2-0.0}^{+0.0+0.2+0.0}$\\

  $B^0 \to K^0(f_2^{\prime}(1525)\to)K^+ K^-$
  &$0.58^{+0.31+0.12+0.05}_{-0.27-0.13-0.01}$
  &$0.13^{+0.12}_{-0.08}\pm0.16$
  &$$\\

  $B^0 \to K^0(f_2(2010)\to)K^+ K^-$
  &$1.09^{+0.57+0.26+0.11}_{-0.48-0.23-0.00}$
  &
  &$$\\
  \hline
  $B^0 \to K_S(f_0(980)\to)K_S K_S$
  &$4.51^{+2.52+1.01+0.34}_{-1.94-1.08-0.00}$
  &$2.7_{-1.2}^{+1.3}\pm0.4\pm1.2$
  &$2.4_{-0.0-0.5-0.0}^{+0.0+0.6+0.0}$\\

  $B^0 \to K_S(f_0(1500)\to)K_S K_S$
  &$0.28^{+0.13+0.05+0.02}_{-0.12-0.08-0.01}$
  &$$
  &$0.15_{-0.00-0.02-0.00}^{+0.00+0.03+0.00}$\\

  $B^0 \to K_S(f_0(1710)\to)K_S K_S$
  &$0.73^{+0.41+0.19+0.06}_{-0.31-0.21-0.00}$
  &$0.50_{-0.24}^{+0.46}\pm0.04\pm0.10$
  &$0.28_{-0.00-0.04-0.00}^{+0.00+0.05+0.00}$\\

 $B^0 \to K_S(f_2^{\prime}(1525)\to)K_S K_S$
 &$0.29^{+0.16+0.06+0.02}_{-0.13-0.07-0.01}$
 &$$
 &$$\\

 $B^0 \to K_S(f_2(2010)\to)K_S K_S$
 &$0.54^{+0.29+0.13+0.06}_{-0.24-0.12-0.00}$
 &$0.54_{-0.20}^{+0.21}\pm0.03\pm0.52$
 &$$\\

 \hline \hline
\end{tabular}
\end{center}
\end{table}

\begin{table}[!t]
\caption{The local $CP$ asymmetries (in $\%$) of various $B\to K^{+/0} (R \to) K^+ K^-/K_S K_S$ decays in PQCD approach. Experimental data are also taken from the BABAR collaboration\cite{Lees:2012kxa}.}
 \label{cp}
\begin{center}
\begin{tabular}{l c c }
 \hline \hline
 \multicolumn{1}{c}{Decay Modes}
 &\multicolumn{1}{c}{PQCD }
 &\multicolumn{1}{c}{EXP\cite{Lees:2012kxa}}   \\
  \hline
  \hline
 $B^+ \to K^+(\phi(1020)\to)K^+ K^-$
 &$5.98^{+5.85+5.07+3.29}_{-2.66-3.08-0.00}$
 &$12.8\pm4.4\pm1.3$ \\
 $B^+ \to K^+(f_0(980)\to)K^+ K^-$
 &$-4.59^{+2.83+1.67+1.20}_{-0.00-0.94-0.66}$
 &$-8\pm8\pm4$ \\
 $B^+ \to K^+(f_0(1500)\to)K^+ K^-$
 &$14.1^{+8.7+1.9+2.5}_{-2.7-1.5-0.0}$
 &   \\
 $B^+ \to K^+(f_0(1710)\to)K^+ K^-$
 &$-0.73^{+4.11+1.89+1.07}_{-0.00-0.00-1.00}$
 & \\
 $B^+ \to K^+(f_2^{\prime}(1525)\to)K^+ K^-$
 &$-10.3^{+3.4+3.1+1.7}_{-0.0-0.2-0.0}$
 &$14\pm10\pm4$  \\
 $B^+ \to K^+(f_2(2010)\to)K^+ K^-$
 &$-9.13^{+5.25+5.28+2.86}_{-0.00-0.10-0.00}$
 &$$ \\
 \hline
 $B^+ \to K^+(f_0(980)\to)K_S K_S$
 &$-0.04^{+2.83+1.67+1.21}_{-0.00-0.94-0.66}$
 &$$    \\

 $B^+ \to K^+ (f_0(1500)\to)K_S K_S$
 &$12.1^{+8.86+2.78+3.23}_{-2.33-1.03-0.00}$
 &$$     \\

 $B^+ \to K^+(f_0(1710)\to)K_S K_S$
 &$-0.07^{+4.07+1.90+1.25}_{-0.00-0.00-0.73}$
 &$$   \\

 $B^+ \to K^+(f_2^{\prime}(1525)\to)K_S K_S$
 &$-10.3^{+3.41+3.06+1.73}_{-0.00-0.16-0.00}$
 &\\

 $B^+ \to K^+(f_2(2010)\to)K_S K_S$
 &$-11.7^{+3.32+2.38+0.48}_{-3.52-2.41-0.00}$
 &\\

 \hline
  $B^0 \to K^0(\phi(1020)\to)K^+ K^-$
  &$0.0$
  &\\

  $B^0 \to K^0(f_0(980)\to)K^+ K^-$
  &$1.05^{+3.29+1.29-1.46}_{-0.00-0.49-0.63}$
  &\\

  $B^0 \to K^0(f_0(1500)\to)K^+ K^-$
  &$-1.42^{+7.54+0.00+1.38}_{-4.99-2.49-0.00}$
  &\\

  $B^0 \to K^0(f_0(1710)\to)K^+ K^-$
  &$1.36^{+3.84+2.40+1.16}_{-0.00-0.00-0.96}$
  & \\

  $B^0 \to K^0(f_2^{\prime}(1525)\to)K^+ K^-$
  &$-2.29^{+2.91+1.23+1.11}_{-1.18-2.11-0.34}$
  & \\

  $B^0 \to K^0(f_2(2010)\to)K^+ K^-$
  &$0.97^{+1.15+0.43+0.00}_{-3.51-2.90-0.79}$
  &\\
  \hline
  $B^0 \to K_S(f_0(980)\to)K_S K_S$
  &$2.10^{+3.29+1.28+1.46}_{-0.00-0.49-0.63}$
  &\\

  $B^0 \to K_S(f_0(1500)\to)K_S K_S$
  &$-1.42^{+7.54+0.00+1.38}_{-4.99-2.49-0.00}$
  &\\

  $B^0 \to K_S(f_0(1710)\to)K_S K_S$
  &$1.36^{+3.84+2.41+1.16}_{-0.00-0.00-0.96}$
  &\\

  $B^0 \to K_S(f_2^{\prime}(1525)\to)K_S K_S$
  &$-2.29^{+2.92+1.24+1.11}_{-1.18-2.11-0.34}$
  &\\

  $B^0 \to K_S(f_2(2010)\to)K_S K_S$
  &$0.97^{+1.16+0.43+0.00}_{-3.50-2.90-0.78}$
  &\\

 \hline \hline
\end{tabular}
\end{center}
\end{table}

\begin{figure}[!htb]
\begin{center}
\includegraphics[scale=0.35]{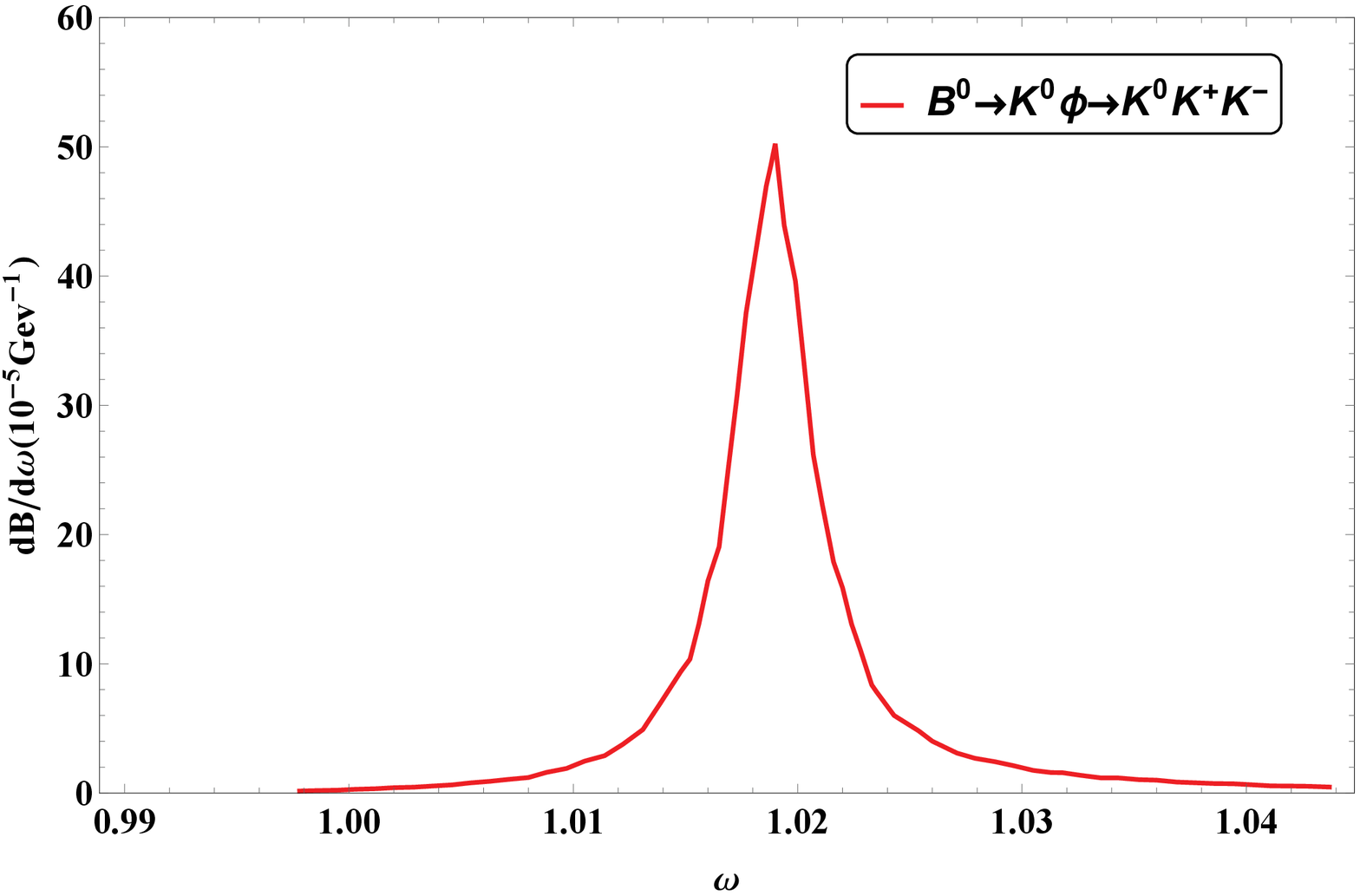}\,\,\,\,
\includegraphics[scale=0.35]{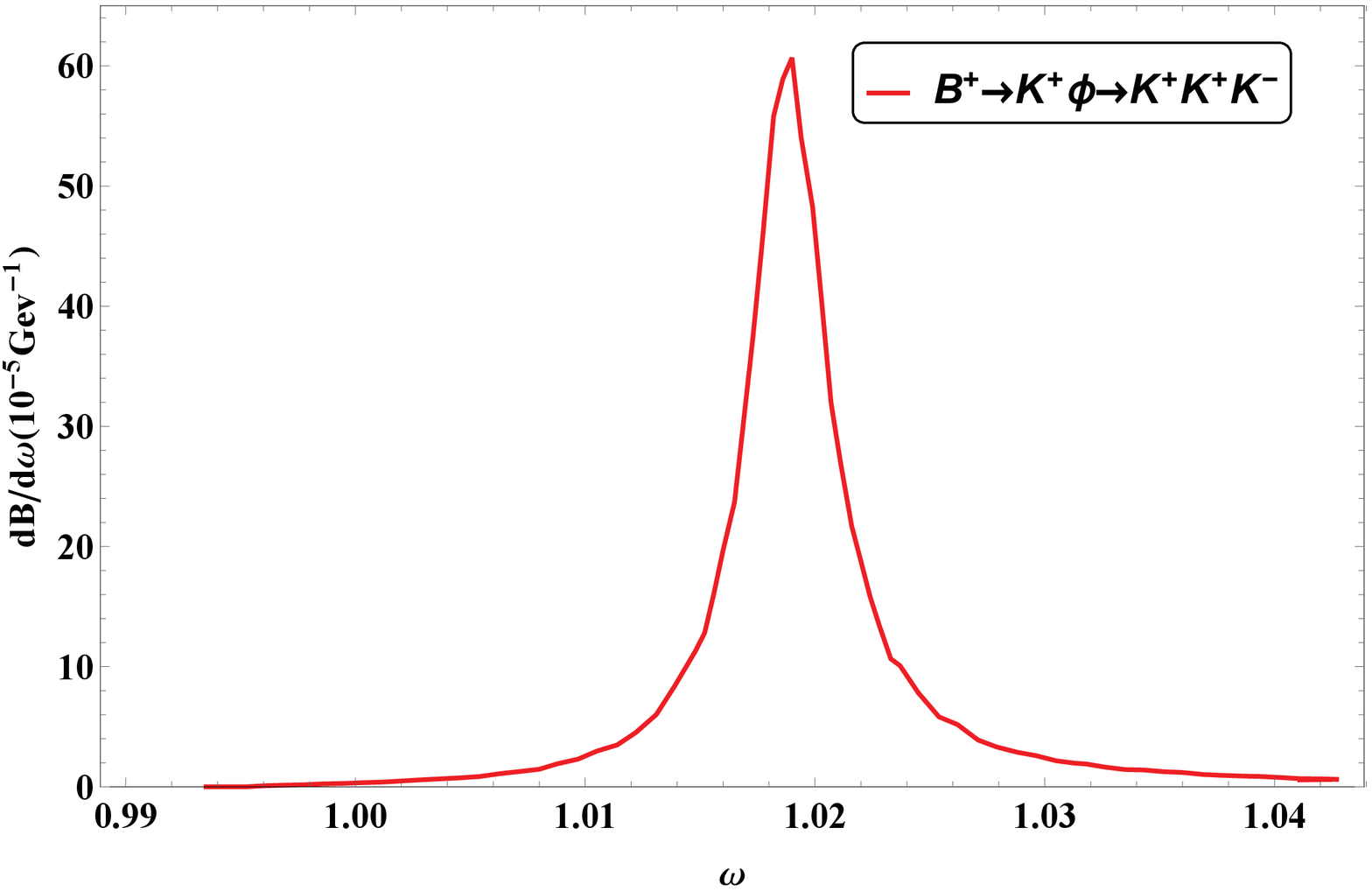}\,\,\,\,
\caption{The $\omega$-dependence of differential branching fractions for the $B\to K\phi \to KKK$ decays.}\label{Fig:2}
\end{center}
\end{figure}
\begin{figure}[!htb]
\begin{center}
\includegraphics[scale=0.5]{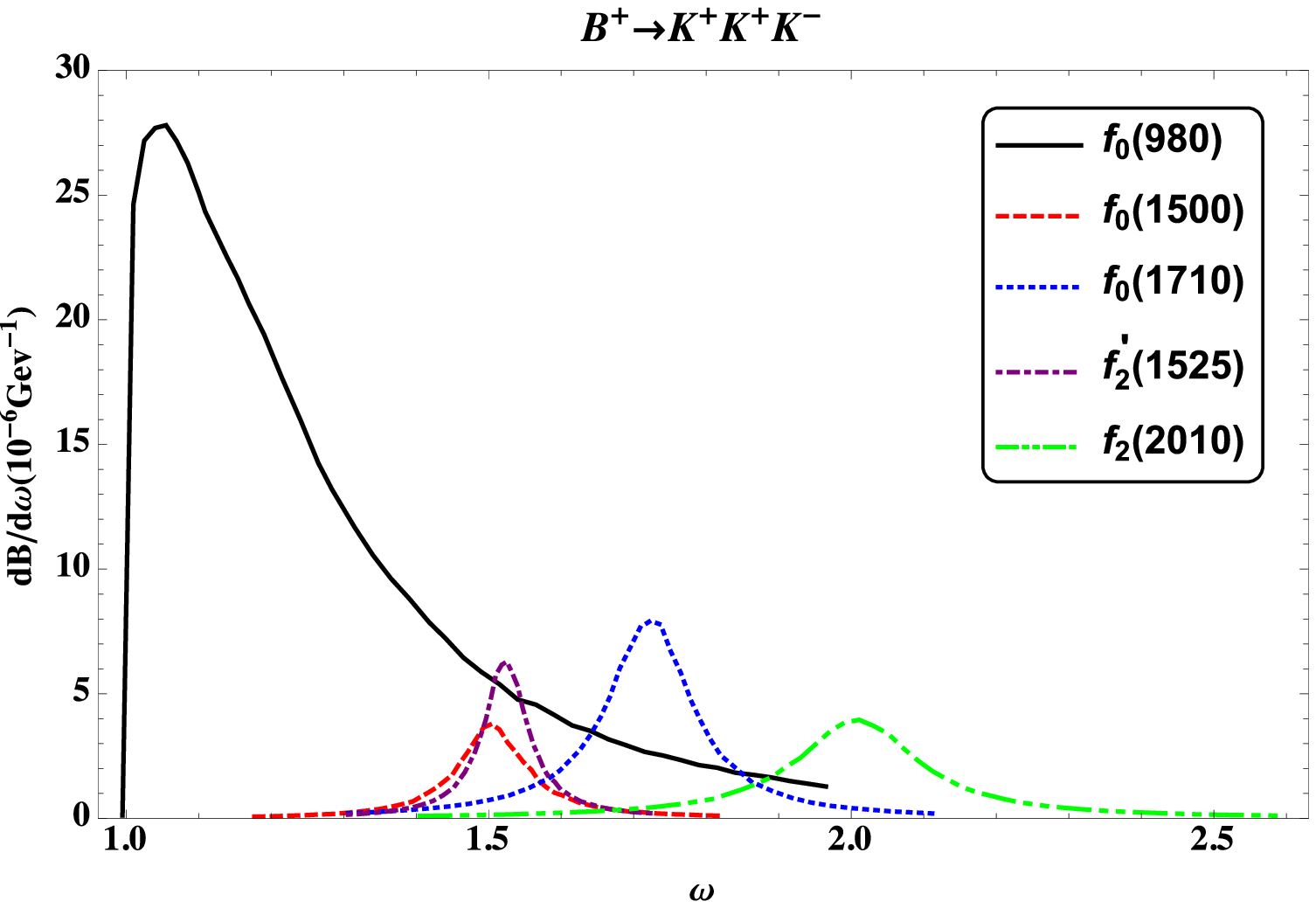}\,\,\,\,
\includegraphics[scale=0.5]{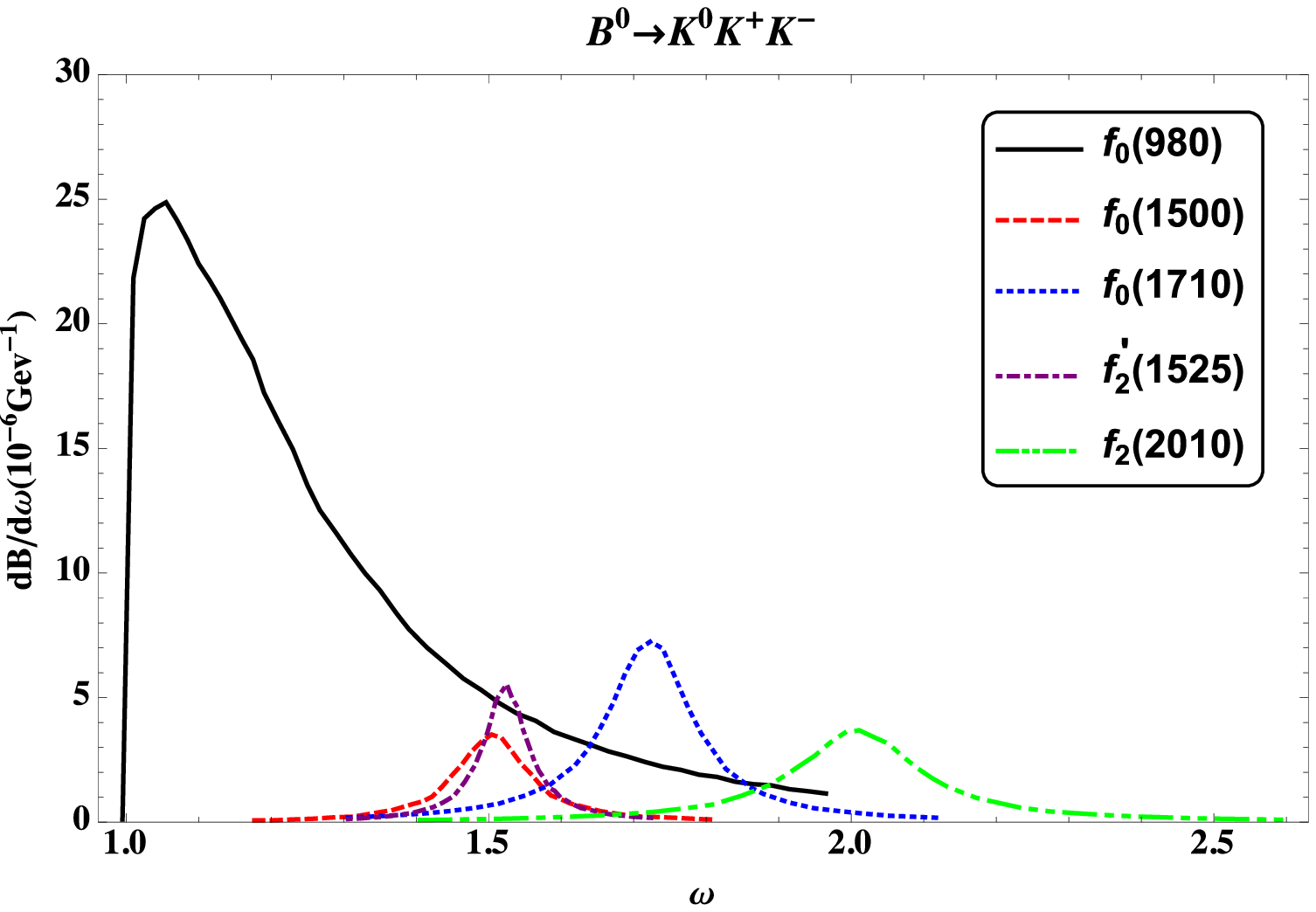}\,\,\,\,\\
\includegraphics[scale=0.5]{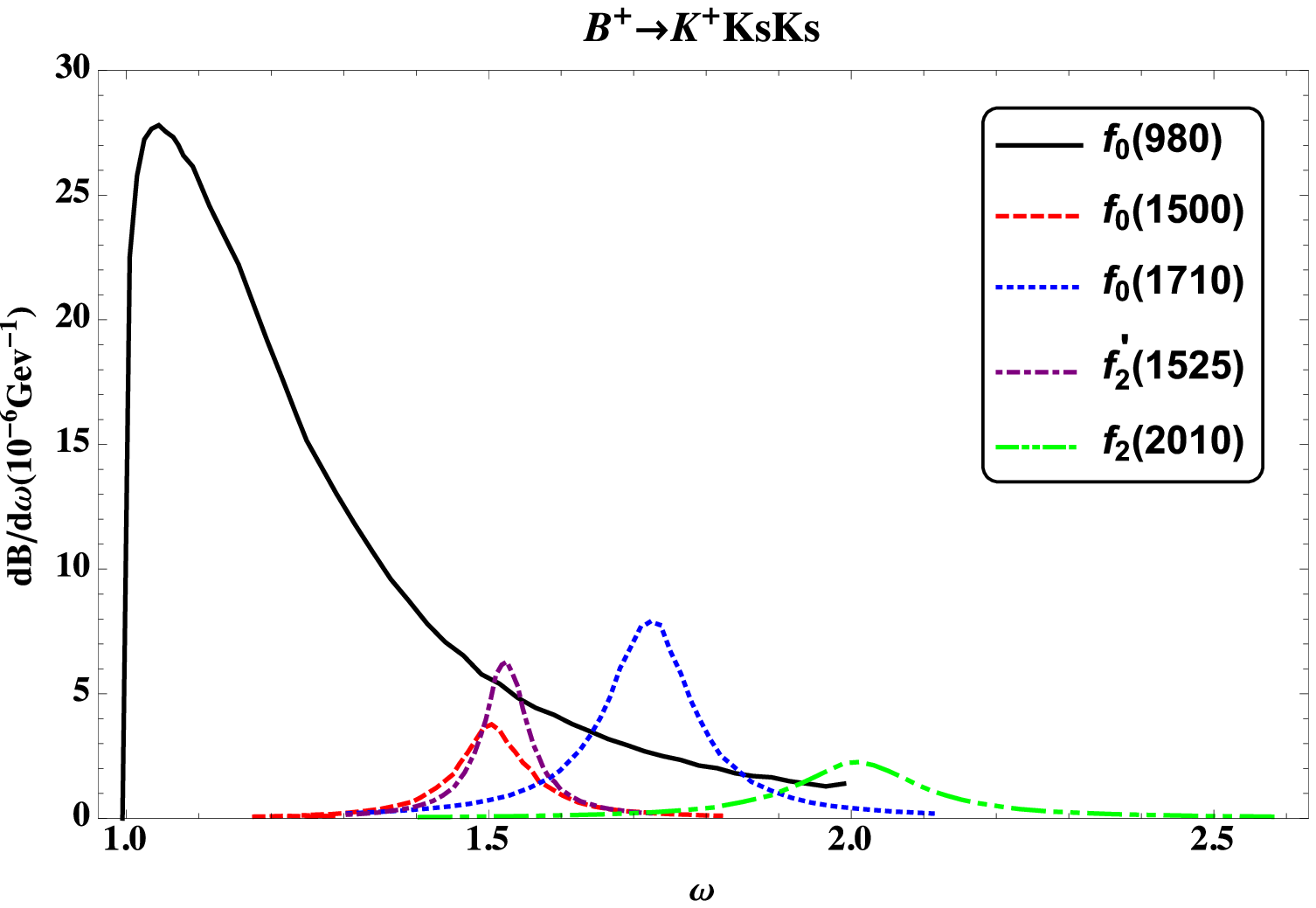}\,\,\,\,
\includegraphics[scale=0.5]{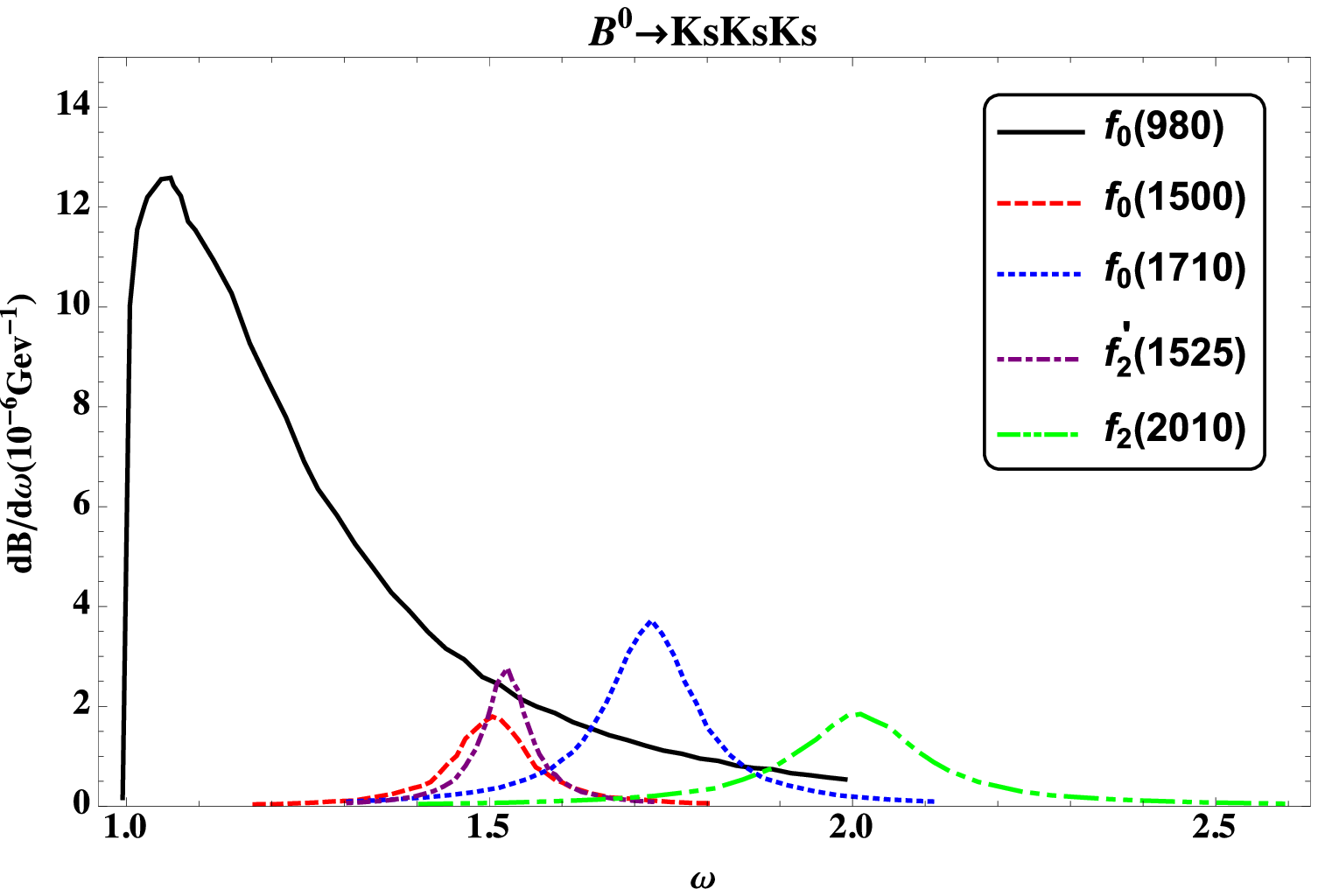}\,\,\,\,\\
\caption{The $\omega$-dependence of differential branching fractions from $f_0(980)$, $f_0(1500)$, $f_0(1710)$, $f_2^\prime(1525)$ and $f_2(2010)$ for the $B\to KKK$ decays.}\label{Fig:3}
\end{center}
\end{figure}

Within the amplitudes presented in Sec.\ref{sec:amplitude} and above parameters, we calculate the $CP$ averaged branching fractions and the direct $CP$ asymmetries for the concerned quasi-two-body decays $B\to KR \to KKK$, and present them in Tables.\ref{br} and \ref{cp}, together with some currently available experimental measurements. For comparison, we also list the results of the factorization approach \cite{Cheng:2013dua}. To be honest, there are many uncertainties in our calculations, and we here mainly consider three kinds of them. The first errors are from nonperturbative inputs, which manifest in the distribution amplitudes of $B$ meson, kaon and kaon-pair. In our calculations, we focus on the $B$ meson decay constant $f_B$ and its shape parameter $\omega_B=0.4\pm0.04 ~{\rm GeV}$, the Gegenbauer moments in the distribution amplitudes of $K$ meson, and the Gegenbauer moments $a_{S(V,T)}$ in the distribution amplitudes of kaon-pair, whose values are varied with a $20\%$ range. It is emphasized that this kind errors are dominant, and they will decrease with the improvement of the experiments and the update of the theoretical understanding. The second kind of errors come from the unknown QCD radiative corrections and the power corrections characterized by varying the $\Lambda_{QCD}=0.25\pm 0.05 ~{\rm GeV}$ and factorization scale $t$ from $0.8t$ to $1.2t$, respectively. The last kind of uncertainties are caused by the CKM matrix elements, and this kind uncertainties are the smallest ones.  For the direct $CP$ asymmetries, it is found from Table.~\ref{cp} that besides the first kind errors, the second kind errors and the third ones also become dominant because they could affect the strong phases and weak phases remarkably. In the experimental side, only few data on these decays with large uncertainties were reported. For decays $B^+ \to K^+ (\phi(1020),f_0(980)) \to K^+ K^+ K^-$, our results can agree with data well. As for $B^+ \to K^+ f_2^{\prime}(1525) \to K^+ K^+ K^-$, although our prediction and experimental data have opposite sign, both of them have large uncertainties. We hope this discrepancy can be settled with theoretical improvement  and high precision measurement in the experiments in future.

From the Table.\ref{br}, one can find that within the uncertainties most of our results are in good agreement with experimental results \cite{Lees:2012kxa,Lees:2011nf} of BaBar, except two decay modes $B^+ \to K^+f_0(1710)\to K^+K_S K_S$ and $B^0 \to K^0f_0(1710)\to K^0K^+ K^-$, which will be discussed in further detail below.  In 2005, Belle also studied the amplitude of the three-body charmless decay $B^+\to K^+K^-K^+$ in detail in ref.\cite{Garmash:2004wa}. For the quasi-two-body decay $B^+\to K^+\phi\to K^+K^+K^-$, Belle measured the branching fraction to be $(4.72\pm0.45\pm0.35^{+0.39}_{-0.22})\times 10^{-6}$, which is consistent with result of BaBar \cite{Lees:2012kxa}. Besides the $f_0(980)$ and $f_2^{\prime}(1525)$ resonances, Belle also analyzed the events of the $\phi(1680)$ and $a_2(1320)$ resonances, but the signals of these two particles are not clear enough to provide any information for theoretical studies. For this reason, we have not taken $\phi(1680)$ and $a_2(1320)$ resonances into account in this present work.

Let us first discuss the $P$-wave contribution in quasi-two-body decays $B\to K\phi \to KK^+K^-$. To study the contribution of the $\phi$ resonance, we show the $K^+K^-$ invariant mass-dependent differential branching fractions for the quasi-two-body decays $B\to K\phi \to KK^+K^-$ in Fig.~\ref{Fig:2}. It is found that the main portion of branching fractions for $B\to K\phi \to KK^+K^-$ comes from the region around the pole mass of the resonant state $\phi$. In 2005, Belle first obtained the branching fraction of $B^+\to K^+\phi$ decay to be $(9.60 \pm0.92 \pm 0.71_{-0.46}^{+0.78})\times 10^{-6}$ \cite{Garmash:2004wa}. Subsequently, in 2012, BaBar also measured that the branching fractions of $B^+\to K^+\phi$ and $B^0\to K^0\phi$ decays are $(9.2\pm 0.4_{-0.5}^{+0.7})\times 10^{-6}$ and $(7.1 \pm 0.6_{-0.3}^{+0.4}) \times  10^{-6}$ \cite{Lees:2012kxa} respectively, which are consistent with the results of Belle. Thus, the averaged branching fractions of $B^+\to K^+\phi$ and $B^0\to K^0\phi$ decays are $(8.8_{-0.6}^{+0.7}) \times 10^{-6}$ and $(7.3\pm 0.7) \times 10^{-6}$ \cite{Tanabashi:2018oca}. Under the narrow-width approximation, the three-body decay and corresponding two-body one satisfy the factorization relation
\begin{eqnarray}
\mathcal{B}(B\to P {\cal R}\to P P_1P_2)=\mathcal{B}(B\to P {\cal R})\times\mathcal{B}({\cal R}\to P_1P_2),
\label{nwa}
\end{eqnarray}
with ${\cal R}$ being the resonance. Based on the decay rate $\mathcal{B}(\phi\to K^+K^-)=(49.2\pm0.5)\%$ \cite{Tanabashi:2018oca}, we use our results in Table.~\ref{br} and obtain that the branching fractions of $B^+\to K^+\phi$ and $B^0\to K^0\phi$ decays are $(7.8^{+3.2}_{-2.2}) \times 10^{-6}$ and $(6.4^{+2.9}_{-2.0})\times 10^{-6}$, which are in agreement with above experimental results with uncertainties. In ref.~\cite{Li:2006jv}, these two-body decays have been investigated within PQCD approach, and our results agree with their results well. Because the process $\phi \to K_SK_S$ violates the Pauli exclusion principle, the quasi-two-body decays $B\to K\phi \to K K_SK_S$ are prohibited strictly.

At this stage, we shall discuss the contributions from $S$-wave particles. In contrast to vector resonance, the quark structure of scalar particles are still quite controversial, especially for the light scalar ones. Although there are many hints that the light scalars are four-quark states, we here still regard $f_0(980)$ as two quark structure. In two-quark picture, many experimental evidences indicate that both $s\bar s$ and $q\bar q$ are involved in the $f_0(980)$, and the mixing form is given by \cite{Cheng:2002ai}
\begin{eqnarray}
|f_0(980)\rangle=|q\bar{q}\rangle\sin\theta+|s\bar{s}\rangle\cos\theta,
\end{eqnarray}
with $q\bar{q}=(u\bar{u}+d\bar{d})/\sqrt{2}$. The value of the mixing angle $\theta$ is not well determined so far, as it varies considerably in different analysis. For example, the fraction between $J/\psi\to f_0(980)\phi$ and $J/\psi\to f_0(980)\omega$ allows the mixing angle to be $(34\pm6)^{\circ}$ and $(146\pm6)^{\circ}$. The analysis of three-body decay $D_s\to \pi^+\pi^-\pi^+$ determines $35^{\circ}<\mid \theta\mid<55^{\circ}$. A value $\theta=(42.14^{+5.8}_{-7.3})^{\circ}$ can be inferred from the ratio between $D_s^+\to f_0(980)\pi^+$ and $D^+\to f_0(980)\pi^+$. The analysis from the light-cone QCD sum rules prefers the values $(27\pm13)^{\circ}$ and $(41\pm 11)^{\circ}$. Therefore, based on the experimental measurements we fix the value of $\theta$ as $40^{\circ}$. It is well known that there are glueball contents in isosinglet scalar mesons $f_0(1370)$, $f_0(1500)$ and $f_0(1710)$. It is commonly accepted that $f_0(1710)$ is dominated by the scalar glueball, while $f_0(1500)$ is an approximately SU(3) octet with negligible glueball component. In view of this, the glueball content of $f_0(1500)$ will be neglected in this work. Moreover, since the study in ref.\cite{Cheng:2006hu} indicates that the scalar glueball decaying to two pseudoscalar mesons are chiral suppressed, we only study the effects of the quark component in $f_0(1710)$ when discussing the effects of $f_0(1710)$ in the decays $B\to KKK$.

The predicted dependencies of the differential branching ratios $d{\cal B}/d\omega$ on the kaon-pair invariant mass $\omega$ are presented for the $S$-wave resonances $f_0(980)$, $f_0(1500)$ and $f_0(1710)$ in the $B\to KKK$ decays in Fig.~\ref{Fig:3}, where the results of $D$-wave particles $f_2^\prime(1525)$ and $f_2(2010)$ are also shown. The different shapes among these individual channels are mainly governed by the corresponding kaon-pair functions and parameters $a_i$ in Eq.~(\ref{isobar}). As expected, the $f_0(980)$ productions are apparently dominant, and they are about ten times larger than $f_0(1710)$ productions. Furthermore, because these particles have large widths, the effects of the tail of $f_0(980)$ are still larger than the effects of $f_0(1500)$. Furthermore, the contributions of  $f_0(1710)$ and  $f_0(1500)$ overlap with each other.  As a result, at the region about $1.5~\rm GeV$, the effects from all $S$-wave resonances are intertwined, and it is very hard for us to disentangle them. Moreover, such entanglements make the $CP$ asymmetries become more complicated than ones of two-body decays.

From Table.~\ref{br} it is seen that for these decays involving $f_0(980)$ resonance our predictions agree with the BaBar measurements well within errors. It should be noted that in our calculations the two-meson wave functions rather than the narrow-width approximation have been used, both resonant and nonresonant effects are all included. If under the narrow-width approximation, we use the averaged experimental measurements \cite{Tanabashi:2018oca} of quasi-two-body decays $B^+ \to K^+f_0(980) \to K^+K^+K^-$ and $B^+\to K^+f_0(980)$ $\to K^+ \pi^+\pi^-$ and obtain the ratio between the $f_0(980)\to K^+K^-$ and $f_0(980) \to \pi^+\pi^-$ as
\begin{eqnarray}
{\cal R}_1\equiv\frac{\mathcal{B}(f_0(980)\to K^+K^-)}{\mathcal{B}(f_0(980) \to \pi^+\pi^-)}
=\frac{\mathcal{B}(B^+ \to K^+f_0(980)\to K^+K^+K^-)}{\mathcal{B}(B^+\to K^+f_0(980)\to K^+\pi^+\pi^-)}
\sim 1.0^{+0.5}_{-0.4}.
\end{eqnarray}
In ref.~\cite{Aubert:2006nu}, using the decays $B \to KK^+K^-$ and $B\to K\pi^+\pi^-$, BaBar measured this ratio to be ${\cal R}_1=0.69\pm0.32$, however it changes to  $0.92\pm0.07$ if the input parameters of $f_0(980)$ were adopted from BES \cite{Ablikim:2004wn}. Meanwhile, BES measured ${\cal R}_1\sim 0.625\pm0.21$ \cite{Ablikim:2004wn} by studying the decays $J/\psi\to \phi f_0(980) \to\phi\pi^+\pi^-$ and $J/\psi \to \phi f_0(980)\to \phi K^+K^-$. In refs.\cite{Ablikim:2004cg, Ablikim:2005kp}, BES also obtained ${\cal R}_1= 0.25^{+0.22}_{-0.20}$ by analyzing the results of the decays $J/\psi\to \gamma\chi_{c0}\to \gamma f_0(980)f_0(980)\to\gamma \pi^+\pi^-K^+K^-$ and $J/\psi \to \gamma\chi_{c0}\to \gamma f_0(980)f_0(980)\to\gamma \pi^+\pi^-\pi^+\pi^-$. By studying the decays $B_s \to J/\psi \pi^+\pi^-$ and $B_s \to J/\psi K^+K^-$, the authors also estimated this ratio to be $0.37_{-0.13}^{+0.23}$ \cite{Rui:2019yxx} within the narrow-width approximation. Overall, it seems that we hardly can reach a reliable and universal ${\cal R}_1$, and even the PDG have not performed the averaged value using the current experimental data. In fact, in multi-body decays where the resonance $f_0(980)$ is involved, it is off-shell when the final states are $K^+K^-$. However, under the narrow-width approximation it is particularly viewed as on-shell when it decays to $\pi^+\pi^-$. So, the narrow-width approximation may be invalid in processes where the resonance $f_0(980)$ decays to $K^+K^-$, and that is the reason why under the narrow-width approximation ${\cal R}_1$ varies so much in different measurements.

Supposing the narrow-width approximation relation is valid in process $B^{0}\to Kf_0(1500) \to KK^+K^-$, we can then obtain the branching fractions of $B\to Kf_0(1500)$ as
 \begin{eqnarray}
 \mathcal{B}(B^{0}\to K^{0}f_0(1500))&=&(13.7\pm6.1)\times 10^{-6},\\
 \mathcal{B}(B^{+}\to K^{+}f_0(1500))&=&(13.9\pm5.8)\times 10^{-6},
 \end{eqnarray}
within the branching fraction of $f_0(1500)\to K^+K^-$ being $4.3\%$. For the decay $B^{0}\to K^{0}f_0(1500)$, our result agree with both experimental data \cite{Tanabashi:2018oca} and previous studies \cite{Wang:2006ria}.  As for the decay $B^{+}\to K^{+}f_0(1500)$, our result is about 3.7 times larger than the averaged experimental data $ (3.7 \pm 2.2)\times 10^{-6}$ \cite{Tanabashi:2018oca}, but  consist with the previous PQCD prediction $10\times10^{-6}$ \cite{Wang:2006ria}. Under the narrow-width approximation we get the ratio
\begin{eqnarray}
\mathcal{R}_2=\frac{\mathcal{B}(f_0(1500)\to K^+K^-)}{\mathcal{B}(f_0(1500)\to \pi^+\pi^-)}=\frac{\mathcal{B}(B\to Kf_0(1500)\to KK^+K^-)}{\mathcal{B}(B \to K f_0(1500)\to K \pi^+\pi^-)}.
\end{eqnarray}
Using the experimental data $\mathcal{B}(f_0(1500)\to K^+K^-)=4.3\%$ and $\mathcal{B}(f_0(1500)\to \pi^+\pi^-)=23.27\%$ \cite{Tanabashi:2018oca}, we can get the fraction $\mathcal{R}_2=0.185$. Thereby, the branching fractions of $B\to Kf_0(1500) \to K \pi^+ \pi^-$ decays are predicted to be
 \begin{eqnarray}
 \mathcal{B}(B^{+}\to K^{+}f_0(1500)\to K^{+}\pi^+\pi^-)&=&(3.24\pm 1.35)\times10^{-6},\\
 \mathcal{B}(B^{0}\to K^{0}f_0(1500)\to K^{0}\pi^+\pi^-)&=&(3.15\pm1.40)\times10^{-6},
\end{eqnarray}
which can be tested in the ongoing LHCb and Belle-II experiments.

Here we present some comments on $f_0(1500)$. Before 2019, the broad structure of $f_X(1500)$ has already been observed in the analysis of $B^0\to K_S K^+K^-$ and $B^{\pm}\to K^{\pm}K^+K^-$ decays by BaBar \cite{Aubert:2006nu, Aubert:2009av} and Belle \cite{Garmash:2004wa,Nakahama:2010nj}, whose possible candidates are the $f_0(1370)$, $f_0(1500)$, $f_2(1525)$ and $f_0(1710)$. In the process $B^{\pm}\to \pi^{\pm}K^+K^-$ BaBar had also found the broad peak around $1.5~\rm GeV$ \cite{Aubert:2007xb}, while no evidence of the $f_X(1500)$ has been seen in decays $B^{\pm}\to\pi^{\pm} K_sK_s$ \cite{Aubert:2008aw} and  $B^0\to K_sK_sK_s$ \cite{Lees:2011nf}. The peak between 1.5 and 1.6 GeV can also be described by the interference between the $f_0(1710)$ and other nonresonant components. So much for that, the vector structure of the $f_X(1500)$ can not be ruled out. Although in $B^+\to\pi^+ K_sK_s$ decay, where the $f_X(1500)$ is referred as the combined contribution from $f_0(1500)$, $f_2(1525)$ and $f_0(1710)$, BaBar provided the corresponding branching fractions with so large uncertainties, therefore the signal may be incredible and should be further confirmed with the larger data sample. We can not assert the observation of process $f_X(1500)\to K_SK_S$ so far. In 2019, LHCb have found a broad peak near 1.5 GeV \cite{Aaij:2019qps} with respect to the vector resonance $\rho(1450)$. Whether the $\rho(1450)$ is the so-called $f_X(1500)$ needs more detailed researches, which will be left in our next work \cite{rho1450}.

In the experiment, the ratio of the $\mathcal{B}(B^{+}\to K^{+}f_0(1710)$$\to$$ K^{+}K^+K^-)$ to $\mathcal{B}(B^{0}\to K^{0}f_0(1710)\to K^{0}K^+K^-)$ is about $1/4$, while it is as large as $1.0$ in our calculation, which is in agreement with results in ref.\cite{Cheng:2013dua}. If we scrutinize these quasi-two-body decays involving the $S$-wave particle $f_0(1710)$, we also find that the branching fractions of $\mathcal{B}(B^{+}\to K^{+}f_0(1710)\to K^{+} K^+K^-)$ and $\mathcal{B}(B^{0}\to K_Sf_0(1710) \to  K_SK_SK_S)$ agree with data well, while the results of $\mathcal{B}(B^{+}\to K^{+}f_0(1710)\to K^{+} K_SK_S)$ and $\mathcal{B}(B^{0}\to K_Sf_0(1710) \to  K^0K^+K^-)$ cannot accommodate the experimental data, though our results are in agreement with the theoretical results \cite{Cheng:2013dua} based on factorization approach. It is noted that there are large uncertainties in both experimental measurements and the theoretical calculations, so the discrepancy between the data and the theoretical results could be clarified with the high precision experimental data and the deeper theoretical understanding of multi-body decays.  What's more, the branching fractions of those decays with $f_0(1500)$ resonance are smaller than these decays with $f_0(1710)$ resonance,  the main reason of which is that the strong coupling constant $g^{f_0(1500)\to KK}=~0.69 {\rm GeV}$ is much smaller than $g^{f_0(1710)\to KK}=1.6~{\rm GeV}$.  Similarly, we also define a ratio  as
\begin{eqnarray}
\mathcal{R}_3=\frac{\mathcal{B}(f_0(1710)\to K^+K^-)}{\mathcal{B}(f_0(1710)\to \pi^+\pi^-)}=
\frac{\mathcal{B}(B\to Kf_0(1710)\to KK^+K^-)}{\mathcal{B}(B\to Kf_0(1710)\to K\pi^+\pi^-)},
\end{eqnarray}
where the second step is based on the narrow-width approximation. Using the averaged value of $\Gamma (f_0(1710) \to \pi \pi)/\Gamma(f_0(1710)\to K\overline{K})=0.23\pm0.05$ \cite{Tanabashi:2018oca}, we then get the ratio as
\begin{eqnarray}
\mathcal{R}_3=\frac{3}{4}\frac{\Gamma(f_0(1710)\to K\overline{K})}{\Gamma(f_0(1710)\to \pi\pi)}=3.26\pm 0.07.
\end{eqnarray}
Based on the above value and our results of $\mathcal{B}(B\to Kf_0(1710)\to KK^+K^-)$,  we can  predict the branching fractions of $B \to K f_0(1710)\to K \pi^+\pi^-$ decays as
\begin{eqnarray}
 \mathcal{B}(B^{+}\to K^{+}f_0(1710)\to K^{+}\pi^+\pi^-)
&=&(5.0^{+3.9}_{-3.4})\times10^{-7},\nonumber\\
 \mathcal{B}(B^{0}\to K^{0}f_0(1710)\to K^{0}\pi^+\pi^-),
&=&(4.5^{+3.9}_{-3.4})\times10^{-7}.
\end{eqnarray}
and these results are expected to be measured in LHCb and Belle-II experiments.

Now, we come to discuss the contributions of the $D$-wave resonances. Also, from Table.~\ref{br}, it is found that our results are consistent with the current BaBar measurements. The predicted dependencies of the differential branching ratios $d{\cal B}/d\omega$ for $f_2^\prime(1525)$ and $f_2(2010)$ are shown in Fig.~\ref{Fig:3}. Unlike $S$-wave, the contributions from these two resonances do not overlap any more because of the narrow width of $f_2^\prime(1525)$.  As we already known, the $K\overline K$ channels are dominant in $f_2^{\prime}(1525)$ decays with fraction $(88.7\pm2.2)\%$ \cite{Tanabashi:2018oca}. Based on the predictions to the three-body decays in present work, we then also obtain the branching fractions of two body $B\to Kf_2^{\prime}(1525)$ decays as
\begin{eqnarray}
\mathcal{B}(B^{+}\to K^{+}f_2^{\prime}(1525))&=&(1.51^{+0.90}_{-0.72})\times10^{-6},\\
\mathcal{B}(B^{0}\to K^{0}f_2^{\prime}(1525))&=&(1.30^{+0.74}_{-0.67})\times10^{-6},
\end{eqnarray}
which are in agreement with previous studies \cite{Zou:2012td}. Because the processes $f_2^{\prime}(1525)\to K\overline{K}$ is kinematically allowed, the narrow width approximation is applicable. So we can use  the fraction $\Gamma(f_2^{\prime}(1525)\to \pi\pi)/\Gamma(f_2^{\prime}(1525)\to K\overline{K})=0.0092\pm0.0018$ \cite{Tanabashi:2018oca} and get the branching fractions of quasi-two-body decays $B\to Kf_2^{\prime}(1525)\to K\pi \pi$ as
\begin{eqnarray}
\mathcal{B}(B^{+}\to K^{+}f_2^{\prime}(1525)\to K^{+}\pi^+\pi^-)&=&(8.4^{+7.5}_{-4.8})\times10^{-9},\\
\mathcal{B}(B^{+}\to K^{+}f_2^{\prime}(1525)\to K^{+}\pi^0\pi^0)&=&(4.2^{+3.7}_{-2.4})\times10^{-9},\\
\mathcal{B}(B^{0}\to K^{0}f_2^{\prime}(1525)\to K^{0}\pi^+\pi^-)&=&(7.1^{+6.2}_{-4.3})\times10^{-9},\\
\mathcal{B}(B^{0}\to K^{0}f_2^{\prime}(1525)\to K^{0}\pi^0\pi^0)&=&(3.6^{+3.1}_{-2.1})\times10^{-9}.
\end{eqnarray}

Lastly, we give some remarks on the $CP$ asymmetries. From the Table.~\ref{cp}, one can find that the predicted $CP$ asymmetries are very small, and are consistent with the current BaBar measurements. As a note, these decays are governed by the $b\to sq\bar q$ transition, which is a flavor-changing neutral-current process and suppressed significantly by the loop contributions in SM. So the small direct $CP$ violations of these decays in SM are reasonable. Any large anomalies observed in experiments may be the signals of the new physics beyond SM.

\section{Summary}\label{sec:summary}
In this work we have investigated the quasi-two-body decays $B\to KR\to KKK$ decays with the PQCD framework with $R$ being the vector, scalar, and tensor resonances. In order to describe the dynamics of two collinear particles, we introduce the wave functions of kaon-pair for different angular momentum. By keeping the transverse momenta, we calculated all possible diagrams at leading order, including the hard spectator diagrams and annihilation ones. Most of our numerical results are well consistent with the current measurements from  BaBar and Belle, and also are in agreement with predictions based on the factorization approach. We note that the narrow-width approximation is invalid in the quasi-two-body decays $B\to Kf_0(980)\to KKK$. For other decays, under the narrow-width approximation we can extract the branching fractions of the corresponding two-body decays involving the intermediate resonant states, such as the $B \to K \phi $ whose branching fractions agree with the current experimental data well. Furthermore, we then predict the corresponding decays $B\to KR\to K\pi^+\pi^-$, which are expected to be measured in the ongoing LHCb and Belle-II experiments. Since these decays are all penguin dominant, the $CP$ asymmetries are all small in the standard model. Large anomalies observed in experiments may be the signals of the new physics beyond SM. We also emphasize that there are a large amount of uncertainties in both experiments and theoretical studies, and  we hope in future a large data samples from LHCb and Belle-II could help us reduce these uncertainties.
\section*{Acknowledgment}
We thank Hsiang-nan Li and Hai-Yang Cheng for helpful discussions. This work was supported in part by the National Natural Science Foundation of China under the Grants No. 11705159, 11975195, 11875033, and 11765012, and by the Natural Science Foundation of Shandong province under the Grant No. ZR2018JL001 and No.ZR2019JQ04. X. Liu is also supported by by the Qing Lan Project of Jiangsu Province under Grant No.~9212218405, and by the Research Fund of Jiangsu Normal University under Grant No.~HB2016004. Zou acknowledge the hospitality of the Institute of Physics, Academia Sinica, where part of the work was done.

{\small
\bibliographystyle{bibstyle}
\bibliography{mybibfile}
}
\end{document}